\documentclass[a4paper,11pt]{article}
\usepackage{jcappub} 
\usepackage{lineno}
\usepackage[dvipsnames]{xcolor}
\usepackage{xfrac}
\usepackage[T1]{fontenc} 
\usepackage{aas_macros}
\usepackage{natbib}
\usepackage{bm}
\usepackage{hyperref}
\usepackage{subcaption}
\usepackage{mathrsfs}

\usepackage{xcolor}
\usepackage{soul}

\arxivnumber{1234.56789} 
\title{Baryonification II: Constraining feedback with X-ray and kinematic Sunyaev-Zel'dovich observations}

\author[a]{Michael Kova\v{c},}
\author[b,a]{Andrina Nicola,}
\author[c]{Jozef Bucko,}
\author[d]{Aurel Schneider,}
\author[a]{Robert Reischke,}
\author[e,f]{Sambit K. Giri,}
\author[g]{Romain Teyssier,}
\author[h,i]{Matthieu Schaller,}
\author[h]{Joop Schaye}

\affiliation[a]{Argelander-Institut für Astronomie, Universität Bonn, Auf dem Hügel 71, D-53121 Bonn, Germany}
\affiliation[b]{Jodrell Bank Centre for Astrophysics, Department of Physics and Astronomy, The University of Manchester, Manchester M13 9PL, UK}
\affiliation[c]{Institute for Particle Physics and Astrophysics, ETH Zurich, Wolfgang Pauli Strasse 27, 8093 Zurich, Switzerland}
\affiliation[d]{Department of Astrophysics, University of Zurich, Winterthurerstrasse 190, 8057 Zurich, Switzerland}
\affiliation[e]{Van Swinderen Institute for Particle Physics and Gravity, University of Groningen, Nijenborgh 4, 9747 AG Groningen, The Netherlands}
\affiliation[f]{Department of Astronomy and Oskar Klein Centre, AlbaNova, Stockholm University, SE-10691 Stockholm, Sweden}
\affiliation[g]{Department of Astrophysical Sciences, Princeton University, Princeton, NJ 08540, USA}
\affiliation[h]{Leiden Observatory, Leiden University, PO Box 9513, 2300 RA Leiden, the Netherlands}
\affiliation[i]{Lorentz Institute for Theoretical Physics, Leiden University, PO box 9506, 2300 RA Leiden, the Netherlands}

\emailAdd{mkovac@astro.uni-bonn.de}

\abstract{Baryonic feedback alters the matter distribution on small and intermediate scales, posing a challenge for precision cosmology. The new, component-wise baryonification (BFC) approach introduced in Ref.~\cite{Schneider2025_code} provides a self-consistent framework to model feedback effects for different observables. In this paper we use this framework to fit kinematic Sunyaev-Zel'dovich (kSZ) observations from the Atacama Cosmology Telescope (ACT) alongside halo X-ray gas fractions from eROSITA, investigating baryonic feedback in a cosmological context. We first show that the kSZ data from ACT is consistent with the gas fractions from eROSITA, both suggesting a feedback model that is stronger than what is assumed in most hydrodynamical simulations. This finding is in contrast to older, pre-eROSITA gas fraction measurements that point towards weaker feedback in tension with the kSZ results. We suspect these discrepancies to be due to selection bias in the pre-eROSITA sample, or differences in halo mass estimation between the two data sets. In a further step, we use the BFC model to predict the baryonic suppression of the matter power spectrum. Based on our combined fit to data from ACT and eROSITA, we find a power spectrum suppression that exceeds the percent-level at modes above $k=0.3-0.6 \,h\,\mathrm{Mpc}^{-1}$, growing to 2-8 percent at $k=1\,h\,\mathrm{Mpc}^{-1}$, and to 20-25 percent at $k=5\,h\,\mathrm{Mpc}^{-1}$, consistent with strong-feedback hydrodynamical simulations. Finally, we compare our best-fitting model to the observed gas density and pressure profiles of massive galaxy clusters from the X-COP sample, finding excellent agreement. These results show that BFC provides a self-consistent picture of feedback across mass- and length scales as well as different cosmological observables, thus making it promising for applications to multiwavelength studies to jointly constrain cosmology and baryonic effects.}

\begin{document}
\maketitle
\flushbottom

\section{Introduction}\label{sec: Introduction}

Upcoming large-scale structure surveys such as Euclid\footnote{\url{https://www.esa.int/Science_Exploration/Space_Science/Euclid}}, the Rubin Observatory Legacy Survey of Space and Time (LSST)\footnote{\url{https://www.lsst.org/}} and the Roman Space Telescope\footnote{\url{https://roman.gsfc.nasa.gov/}.} promise to push cosmology into a high-precision regime, but their ability to probe small scales is fundamentally limited by uncertainties in baryonic feedback. Energetic processes caused by active galactic nuclei (AGN), supernova explosions and stellar winds redistribute gas within and around halos, modifying the total matter distribution in the Universe and suppressing power in the nonlinear regime~\cite{Rudd_pkSuppression_2008, VanDaalen2011, Chisari_Pk2019, VanDaalen_Pk2020, Schaye2023_flamingo}. These effects introduce significant modeling challenges for weak lensing and clustering analyses, which rely on an accurate understanding of matter clustering across a wide range of scales. Without a robust strategy to account for baryonic physics, cosmological constraints from these surveys risk being systematically biased or overly conservative~\cite{Semboloni_2011}.

One strategy to mitigate these uncertainties is to apply scale cuts and discard small-scale information (e.g.~\cite{Amon_DES_scalecut_2022, Secco_DES_scalecut_2022}). However, this inevitably reduces the constraining power of the surveys. To unlock the full potential of small-scale cosmology, it is essential to accurately model and marginalize over baryonic effects.

Hydrodynamical simulations provide an invaluable window into the complex interplay between baryons and dark matter, revealing how feedback alters halo profiles and matter statistics (\cite{Schaye2023_flamingo, Springel_TNG_2018, McCarthy_BAHAMAS_2017, Dolag_Magneticum_2016, LeBrun_CosmoOwls_2014, Dubois_HorizonAGN_2014}). However, their application in cosmological analyses is limited by their high computational cost and sensitivity to subgrid modeling assumptions. Since the physical processes governing feedback are not fully resolved, simulations rely on phenomenological subgrid prescriptions. These are often tuned to match specific observables, such as cluster gas fractions, which limits their predictive power in new regimes.

A range of strategies have been developed to mitigate the impact of poorly understood small-scale physics in large-scale structure analyses. One line of work introduces theoretical error models to marginalize over unresolved scales (e.g.,~\cite{Baldauf_theoError_2016, Sprenger_theoError_2019, Moreira_theoError_2021, Maraio_theoError_2025}). Alternative approaches seek to incorporate baryonic physics into theoretical predictions. These include halo-based prescriptions (e.g.~\cite{Mead_HMCode_2020, Semboloni_2013, Debackere_2020}), baryonification methods that modify particle distributions in gravity-only simulations~\cite{Schneider2015_code, Schneider2019_code, Arico_bcm_2024, Anbajagane_bfc_2024, Schneider2025_code}, and scaling the difference between linear and non-linear gravity only power spectra~\cite{Amon_Efstathiou_2022, Schaller_Schaye_2025}. Complementary efforts include principal component analysis (PCA) based approaches (e.g.~\cite{Huterer_PCA_2003, Clarkson_PCA_2010, Sharma_PCA_2022}), painting algorithms to populate halos with synthetic baryonic fields~\cite{Agarwal_painting_2018, Osato_Nagai_2023}, or machine learning emulators trained on hydrodynamical simulations~\cite{Troester_ML_2019}.

The baryonification approach is based on displacing particles in dark-matter-only simulations according to analytic, physically motivated gas and stellar profiles in order to mimic the effects of baryons. This enables fast and flexible predictions of baryonic effects across different cosmologies and feedback scenarios. Compared to hydrodynamical simulations, baryonification encapsulates baryonic uncertainties in a low-dimensional, physically interpretable parameter space.

In particular, the baryonification code (BFC) introduced in~\cite{Schneider2015_code, Schneider2019_code} has been shown to reproduce the matter power spectrum of a wide range of hydrodynamical simulations~\cite{Giri2021_BCemu}. It has also been successfully applied to model weak lensing observables and combinations with X-ray and SZ data~\cite{Schneider2022_WL_xray_kSZ,Bigwood2024}, making it a valuable tool for cosmological analyses.

Importantly, constraining baryonic models requires observational probes that are sensitive to the distribution of baryons in and around dark matter halos. While stars contribute a smaller fraction, the dominant baryonic component is hot, ionized gas, which is accessible through multi-wavelength observations. In particular, the kinematic Sunyaev-Zel'dovich (kSZ) effect provides a direct probe of the line-of-sight momentum of ionized gas, while X-ray measurements trace its thermal state and spatial distribution in galaxy groups and clusters. Recent studies have demonstrated the power of combining these complementary observables to break degeneracies between baryonic and cosmological parameters~\cite{Schneider2022_WL_xray_kSZ, Bigwood2024, Grandis2023, McCarthy_kSZ2024}.

In this work, we build upon the updated BFC model introduced and validated in Paper I~\cite{Schneider2025_code}, where it was shown to reproduce the matter power spectrum across a variety of feedback scenarios and redshifts by fitting to the ratios of gas and stellar mass profiles in the hydrodynamical simulations FLAMINGO~\cite{Schaye2023_flamingo, Kugel_2023} and IllustrisTNG~\cite{Springel_TNG_2018}.

Here, we apply the updated BFC model to jointly fit the kSZ signal from the Atacama Cosmology Telescope (ACT)~\cite{Schaan2021_sz} and X-ray gas fraction measurements from both the recent eROSITA observations~\cite{Popesso2024_fgas} and a compilation of pre-eROSITA datasets~\cite{Lovisari2015_fgas, Gonzalez2013_fgas, Sun2009_fgas}. We focus on modeling the hot gas component, which dominates the baryonic mass budget and is directly probed by these observables. Unlike previous studies, we do not calibrate the model against hydrodynamical simulations but instead constrain it entirely using observational data.

Our results show that the combination of kSZ observations from ACT and X-ray gas fractions from eROSITA yields robust and self-consistent constraints on baryonic feedback when modeled with the updated component-wise BFC framework. The two datasets — each probing different mass, redshift, and radial regimes — are mutually consistent and both favor a strong-feedback scenario. However, feedback manifests differently for the two samples: the lower-mass kSZ groups exhibit a more pronounced suppression, while the higher-mass eROSITA clusters show a comparatively milder effect, consistent with the findings in Ref.~\cite{Lucie-Smith_2025}. This is in contrast to earlier pre-eROSITA measurements, which suggest weaker feedback and are in tension with the kSZ signal. Using the best-fit model, we predict a suppression of the matter power spectrum that exceeds the percent level at $k>0.3\;h/\mathrm{Mpc}$, reaching up to 25\% at $k=5\;h/\mathrm{Mpc}$, in line with strong-feedback hydrodynamical simulations, such as the $f_{\rm gas}-8\sigma$ FLAMINGO simulation. Furthermore, the same model successfully reproduces the gas density and pressure profiles of massive clusters in the X-COP sample, reinforcing its applicability across mass scales and observables. These findings highlight the potential of baryonification-based approaches for jointly constraining cosmology and baryonic physics in upcoming multi-wavelength surveys.

This paper is organized as follows: In Section~\ref{sec: Theory}, we present the theoretical framework underlying the gas and kSZ modeling. In Section~\ref{sec: hydro sims} we give an overview of the hydrodynamical simulations used in this work, and in Section~\ref{sec: Obs data} we describe the observational data. Section~\ref{sec: Inference} outlines the inference pipeline. Our results are presented in Section~\ref{sec: results}, and we conclude with a summary and outlook in Section~\ref{sec: conclusion}.\\

Throughout this work, we adopt the fiducial cosmology used in the FLAMINGO simulations, assuming a flat universe with massless neutrinos with parameters $\Omega_\mathrm{m}=0.306$, $\Omega_\mathrm{b}=0.0486$, $\sigma_8=0.807$, $H_0=68.1\mathrm{\:km\:s^{-1}\:Mpc^{-1}}$, and $n_\mathrm{s}=0.967$. Here $\Omega_\mathrm{m}$ is the fractional matter density today, $\Omega_\mathrm{b}$ denotes the fractional baryon density today, $\sigma_8$ is the r.m.s. of linear matter fluctuations in
spheres of comoving radius $8 h^{-1}$ Mpc, $H_0$ denotes the Hubble parameter, and $n_\mathrm{s}$ is the scalar spectral index.

\section{Theory}\label{sec: Theory}
In this section, we begin with a brief summary of Paper I~\cite{Schneider2025_code}, which introduced the updated BFC model. We then describe how this model is used to simulate cosmological observables, in particular the kSZ signal and X-ray gas fractions.

\subsection{Baryonification Model}\label{subsec: baryonification}

The baryonification model is based on post-processing gravity-only $N$-body simulations to incorporate baryonic feedback effects due to gas and stars. By displacing particles around halo centers according to empirical halo profiles, the model reproduces the impact of baryonic physics on the large-scale structure. We introduce several key advancements in this updated model, described in detail by Ref.~\cite{Schneider2025_code}, including independent treatments of dark matter, gas, and stellar components, and a model for the gas pressure. Throughout this paper, we define the halo mass, $M_{200}$, as the mass enclosed within a radius, centered on the group or cluster, where the mean density equals 200 times the critical density of the Universe unless stated otherwise. 

\subsubsection{Overview of the Method}
In order to separately model all matter components of the Universe, we duplicate each particle in the original gravity-only simulation, creating two sets: one representing dark matter and the other baryonic matter. The masses of these particles are renormalized to ensure the simulation box maintains the correct cosmic baryon and dark matter fractions. Initially, these particles share the same positions, but their displacements differ, reflecting the influence of baryonic physics. 
\begin{align}
    d_{\mathrm{dm}}(\pmb{r}_\mathrm{i,dm}|M_{200},c_{200},\pmb{\theta}_\mathrm{bfc}) = \pmb{r}_\mathrm{f,dm}(M_{\mathrm{dm}}) - \pmb{r}_\mathrm{i,dm}(M_{\mathrm{dm}})\,, \\
    d_{\mathrm{bar}}(\pmb{r}_\mathrm{i,bar}|M_{200},c_{200},\pmb{\theta}_\mathrm{bfc}) = \pmb{r}_\mathrm{f,bar}(M_{\mathrm{bar}}) - \pmb{r}_\mathrm{i,bar}(M_{\mathrm{bar}})\,,
\end{align}
where $\pmb{r}_{i}$ and $\pmb{r}_{f}$ are the inverted cumulative mass profiles of the initial (gravity-only) and final (baryonified) states, $M_{200}$ is the halo mass\footnote{$M_{200}$ and $R_{200}$ are defined so that they correspond to an enclosed overdensity equal to 200 times the critical density of the Universe.}, $c_{200}$ is the halo concentration, and $\pmb{\theta}_\mathrm{bfc}$ represents BFC model parameters. This ensures that we recover the baryonified profiles after displacement. The cumulative mass profiles are given by:
\begin{equation}
    M_{\mathrm{i,f}}(r) = \int_0^r \mathrm{d}s\, s^2 \rho_{\mathrm{i},\mathrm{f}}(s)\ ,
    \label{eq: cummulative_mass}
\end{equation}
where $r$ is the radial distance from the center of the halo.
Particles are displaced radially around halo centers, with their trajectories determined by the density profiles of both the initial and final states.

\subsubsection{Density Profiles}
We model the density profiles for dark matter and baryonic components by combining contributions from the one-halo term (initially described by an NFW profile) and a two-halo (background) term representing the large-scale environment. Initially, both dark matter and baryonic profiles share the same functional form, differing only by their cosmic fractions:
\begin{equation}\label{eq: rho1h+rho2h}
\begin{split}
    \rho_\mathrm{i}(r) &= f_{\rm dm}[\rho_{\rm nfw}(r) + \rho_{\rm 2h}(r)], \quad \rho_\mathrm{f}(r) = \rho_{\rm dm}(r) + f_{\rm dm}\rho_{\rm 2h}(r)\;, \\
    \rho_\mathrm{i}(r) &= f_{\rm bar}[\rho_{\rm nfw}(r) + \rho_{\rm 2h}(r)], \quad \rho_\mathrm{f}(r) = \rho_{\rm gas}(r) + \rho_{\rm stars}(r) + f_{\rm bar}\rho_{\rm 2h}(r)\,,
\end{split}
\end{equation}
where $\rho_{\rm nfw}$ denotes the truncated Navarro-Frenk-White (NFW) profile~\cite{Navarro:1997, Baltz_trNFW_2009, Oguri_trNFW_2011}, and $f_{\rm dm} = \Omega_{\rm dm}/\Omega_\mathrm{m}$ and $f_{\rm bar} = \Omega_{\rm b}/\Omega_\mathrm{m}$ define the cosmic dark matter and baryon fractions, respectively. 
We further subdivide the gas and stellar density into different components as $\rho_{\rm gas} = \rho_{\rm hga} + \rho_{\rm iga}$, $\rho_{\rm stars} = \rho_{\rm cga} + \rho_{\rm sga}$, where the subscripts {$\rm hga, iga, cga, sga$} represent hot gas, cold gas, central galaxies, and satellite galaxies, the four baryonic components considered in our model. The different matter profiles are parameterized as follows:

\paragraph{NFW Profile}
Following Ref.~\cite{Baltz_trNFW_2009, Oguri_trNFW_2011} the dark matter halo is initially described by a truncated NFW profile:
\begin{equation}\label{eq: truncated NFW}
    \rho_{\rm nfw}(r) = \frac{\rho_{\rm nfw,0}}{x_\mathrm{s}(1 + x_\mathrm{s})^2}\frac{1}{(1 + x_\mathrm{t}^2)^2}\;,
\end{equation}
where $x_\mathrm{s} = r/r_\mathrm{s}$ and $x_\mathrm{t} = r/r_\mathrm{t}$. The scale radius $r_\mathrm{s}$ is defined by the halo concentration $c_{200} = r_{200}/r_\mathrm{s}$, and the truncation radius is given by $r_\mathrm{t} = \varepsilon r_{200}$, with $\varepsilon = \varepsilon_0 + \varepsilon_1 \nu$ and $\nu = \delta_\mathrm{c}/\sigma(M)$ being the peak height of the halo with $\delta_\mathrm{c}=1.686$. We fix $\varepsilon_0=4.0$ and $\varepsilon_1=0.5$, which provides a good fit to simulation data (see Ref.~\cite{Diemer_eps_2014}). The normalization $\rho_{\rm nfw,0}$ is set such that the total mass equals $M_{200}$ and is proportional to $1/\int y_{\rm nfw}(r)$.

\paragraph{Gas Profile}
The gas profile is a key component of the baryonification model, as it provides the main link between observable signals (such as X-ray and SZ measurements) and the underlying theoretical description of baryonic matter in halos. It is typically divided into hot and cold components, reflecting the dominant baryonic phases. We focus here on the hot gas component, which dominates the baryonic budget in group- and cluster-scale halos and is particularly sensitive to AGN feedback. Its accurate modeling is therefore essential for interpreting current observations. For further details on the functional modeling of the cold gas component, we refer the reader to Ref.~\cite{Schneider2025_code}. 

\paragraph{Hot Gas Component}
The hot gas component dominates the baryonic content of massive halos and plays a progressively smaller role in lower-mass systems, where cold gas and stars become more significant~\cite{Dave_hgas_2020}. In our model, the hot gas is described by a smoothly declining density profile to account for the extended distribution of the hot gas:
\begin{equation}\label{eq: rho_hga equation}
    \rho_{\rm hga}(r) = \rho_{\rm hga,0}\left[1 + x_{\mathrm{c}}^\alpha\right]^{-\beta(M_{200})/\alpha}\left[1 + x_{\mathrm{t}}^\gamma\right]^{-\delta/\gamma},
\end{equation}
where $x_{\mathrm{c}} = r/r_{\mathrm{c}}$ and $x_{\mathrm{t}} = r/r_{\mathrm{t}}$. Here, $r_{\mathrm{c}} = \theta_{\mathrm{c}} r_{200}$ is the core radius, $r_{\mathrm{t}} = \varepsilon r_{200}$ is the truncation radius, and $\beta(M_{200})$ determines the slope of the profile, which varies with halo mass:
\begin{equation}\label{eq: rho_hga beta}
    \beta(M_{200}) = \frac{3(M_{200}/M_{\mathrm{c}})^\mu}{1 + (M_{200}/M_{\mathrm{c}})^\mu}\;.
\end{equation}
where $M_\mathrm{c}$ and $\mu$ are free parameters controlling the pivot mass and transition steepness, and the parameters $\alpha$, $\delta$, and $\gamma$ in Eq. \eqref{eq: rho_hga equation} define the inner and outer slopes and transition region. This mass dependence allows the model to capture the effects of AGN feedback, which tends to flatten gas density profiles in lower-mass halos by expelling gas from central regions. The normalization $\rho_{\rm hga,0}$ is set to conserve the total hot gas mass in the model.

\paragraph{Stellar Profiles}
The stellar distribution can be subdivided into a central galaxy and satellite galaxies. We model the central galaxy profile to be concentrated in the halo center, i.e.:
\begin{equation}\label{eq: cga}
    \rho_{\rm cga}(r) = \frac{f_{\rm cga}M_{\rm tot}}{4\pi R_\mathrm{h}}\frac{1}{r^2}\exp(-r/R_\mathrm{h})\;,
\end{equation}
where $R_{\mathrm{h}} = 0.03\,r_{200}$ is the galactic scale radius, $f_{\rm cga}$ the central galaxy fraction, and $M_{\rm tot}$ is obtained by integrating the truncated NFW profile to infinity.

Stars not part of the central galaxy, primarily satellite galaxies, are assumed to follow the shape of the dark matter distribution:
\begin{equation}
    \rho_{\rm sga}(r) = \frac{f_{\rm sga}}{4\pi r^2}\frac{\mathrm{d}}{\mathrm{d}r}M_{\rm nfw}(\xi r)\;,
\end{equation}
where $\xi$ encodes the effective contraction or expansion of the dark matter due to baryonic processes. A detailed description of the dark matter back-reaction model, including the parametrization of $\xi$ and its dependence on the baryonic components, is provided in Ref.~\cite{Schneider2025_code}.

\subsubsection{Fractions}
In order to fully specify our model, we also need expressions for the mass fractions of the different components, which are defined at $r=\infty$. Following Ref.~\cite{Moster_fstar_2012} the stellar fraction and its division into central and satellite components are parameterized by
\begin{align}\label{eq: stellar fraction equation}
    f_{\rm star} = N_{\rm star} \left(\left( \frac{M_{200}}{M_{\rm star}} \right)^{-\zeta_\mathrm{star}} + \left( \frac{M_{200}}{M_{\rm star}} \right)^{-\eta}\right)^{-1}, \\ 
    f_{\rm cga} = N_{\rm star} \left(\left( \frac{M_{200}}{M_{\rm star}} \right)^{-\zeta_\mathrm{star}} + \left( \frac{M_{200}}{M_{\rm star}} \right)^{-\eta-d\eta}\right)^{-1}\;,
\end{align}
where the fraction of stars in satellite galaxies is simply given by $f_{\rm sga} = f_{\rm star} - f_{\rm cga}$. Here, the parameter $M_{\rm star}$ is fixed to $2.5 \times 10^{11} M_{\odot}/h$ and $\zeta_\mathrm{star}=1.376$ (see Ref.~\cite{Moster_fstar_2012}), while $\eta$, $d\eta$, and $N_{\rm star}$ are free model parameters. The power-law dependence is motivated by abundance matching results and provides a good fit for massive halos ($M_{200} > 10^{12} M_{\odot}/h$) (see Ref.~\cite{Schneider2019_code}).
The fraction of the cold, inner gas is assumed to be proportional to the central galaxy fraction:
\begin{equation}\label{eq: ciga}
    f_{\rm iga} = c_{\rm iga} f_{\rm cga}\;,
\end{equation}
where $c_{\rm iga}$ is a free model parameter that controls the relative contribution of inner gas to the total baryonic content.
With the stellar and inner gas fractions fixed, the remaining hot gas fraction follows as:
\begin{equation}\label{eq: fgas definition}
    f_{\rm gas} = f_{\rm bar} - f_{\rm star} - f_{\rm iga}\;,
\end{equation}
where the total baryon fraction is given by $f_{\rm bar} =\Omega_\mathrm{b}/\Omega_\mathrm{m}$, i.e. the ratio between baryon ($\Omega_\mathrm{b}$) and total matter abundance ($\Omega_\mathrm{m}$) of the Universe.

\subsubsection{Pressure}\label{subsec: pressure}
Assuming that galaxy clusters and groups are in hydrostatic equilibrium, we can obtain the total halo pressure profile from the gas density and total mass profiles through:
\begin{equation}\label{eq: hydrostatic equil.}
    \frac{\mathrm{d}P_\mathrm{tot}}{\mathrm{d}r}=-\rho_\mathrm{gas}\frac{\mathrm{d}\Phi}{\mathrm{d}r}=-\rho_\mathrm{gas}(r)G\frac{M(r)}{r^2}\;.
\end{equation}
The assumption of hydrostatic equilibrium primarily holds for spherically symmetric, virialized, stable halos that are not undergoing mergers or significant accretion, and becomes increasingly inaccurate at low halo masses (see e.g. Ref.~\cite{Oppenheimer:2018, Braspenning_2024}). Additionally, the total pressure can have significant non-thermal contributions, due to effects such as bulk motions and turbulence from gas accretion or mergers. This becomes particularly pronounced at large halo radii.
In order to account for the latter, we define the total pressure as $P_\mathrm{tot}=P_\mathrm{nth}+P_\mathrm{th}$, where $P_\mathrm{nth}$ represents the non-thermal and $P_\mathrm{th}$ the thermal component of the pressure. Following the work of Ref.~\cite{Shaw2010_pnth} the fraction of non-thermal to thermal pressure can be modeled as
\begin{equation}\label{eq: Pnth}
    \frac{P_\mathrm{nt}}{P_\mathrm{tot}}(z)=\alpha(z)\left(\frac{r}{R_{500}}\right)^{n_\mathrm{nt}}\;,
\end{equation}
where $n_\mathrm{nt}$ parametrizes its radial dependence and $\alpha(z)$ the redshift dependence. We assume that the non-thermal pressure varies with redshift as $\alpha(z)=\alpha_\mathrm{0,nt}f(z)$, where $\alpha_\mathrm{0,nt}$ is the mean non-thermal to total pressure ratio at $z=0$ and $R_{500}$ and $f(z)$ is an increasing or decreasing function of redshift given by:
\begin{equation}\label{eq: pressure profile f(z)}
    f(z)=\mathrm{min}\left[(1+z)^{\beta},\;(f_\mathrm{max}-1)\;\mathrm{tanh}(\beta z)+1\right]\;,
\end{equation}
with $f_\mathrm{max}=4^{-n_\mathrm{nt}}/\alpha_0$ and $\beta$ a free parameter regulating the evolution rate. Following Ref.~\cite{Shaw2010_pnth} $\beta$ is fixed to $\beta=0.5$. The subdivision of the total pressure into thermal and non-thermal components is only valid as long as the non-thermal pressure is smaller than the total one. Therefore, the non-thermal pressure is set equal to the total pressure as soon as Eq. \eqref{eq: Pnth} reaches unity. With this, we compute the thermal pressure from the given non-thermal pressure and the total pressure given by Eq. \eqref{eq: hydrostatic equil.} by a simple subtraction. \\

In Tab.~\ref{tab: parameters}, we summarize all parameters used in the model, including their fixed values where applicable.

\begin{table}[h!]
\renewcommand{\arraystretch}{1.5} 
\setlength{\tabcolsep}{8pt} 
\begin{center}
\begin{tabular}{ccp{8cm}c} \hline\hline
\textbf{Parameter} & \textbf{Prior} & \textbf{Description} & \textbf{Fixed value} \\ \hline          
$\theta_{\mathrm{co}}$ & U[0.001, 0.5] & Core radius of the gas profile relative to $r_{200}$ (Eq.~\ref{eq: rho_hga equation}) & 0.3 \\
$\log_{10}\: M_\mathrm{c}$ &  U[11.0, 15.0] & Pivot mass for the mass dependent slope of the gas profile (Eq.~\ref{eq: rho_hga equation} \& Eq.~\ref{eq: rho_hga beta}) & - \\
$\mu$ & U[0.0, 2.0] & Transition steepness of mass dependent gas parameter (Eq.~\ref{eq: rho_hga equation} \& Eq.~\ref{eq: rho_hga beta}) & - \\
$\delta$ & U[1.0, 11.0] & Characterizes the transition from the inner to the outer part of the gas profile (Eq.~\ref{eq: rho_hga equation}) & - \\
$\eta$ & U[0.0, 0.5] & Total stellar fraction within a halo (Eq.~\ref{eq: stellar fraction equation}) & 0.1 \\     
$d\eta$ & U[0.0, 0.5] & Fraction of central galaxies with respect to the total stellar fraction (Eq.~\ref{eq: stellar fraction equation}) & 0.22 \\
$N_{\mathrm{star}}$ & U[0.0, 0.05] & Stellar fraction amplitude (Eq.~\ref{eq: stellar fraction equation}) & 0.030 \\
$c_\mathrm{iga}$ & U[0.0, 1.0] & Cold gas fraction (see Eq.~\ref{eq: ciga}) & 0.1 \\
$b_{\mathrm{hse}}$ & $\mathcal{N}$(0.26, 0.07) & Hydrostatic mass bias & - \\
$\alpha_{\mathrm{0,nt}}$ & - & Non-thermal pressure amplitude (see Eq.~\ref{eq: Pnth}) & 0.1 \\\hline\hline
\end{tabular}
\end{center}
\caption{Summary of the BFC model parameters, their prior distributions, brief descriptions, and fixed values where applicable.}
\label{tab: parameters}
\end{table}

\subsection{Kinematic Sunyaev-Zel'dovich effect}\label{subsec: sz effects}

The kinematic Sunyaev-Zel'dovich effect is a secondary cosmic microwave background (CMB) anisotropy. It is caused by inverse Compton scattering of CMB photons with free electrons in galaxies and clusters with nonzero bulk motion. This inverse scattering shifts the thermodynamic temperature of the CMB according to:
\begin{equation}\label{eq: kSZ}
    \frac{\Delta T_\mathrm{kSZ}}{T_\mathrm{CMB}} = \frac{\sigma_\mathrm{T}}{c} \int_\mathrm{los} \mathrm{d}l\,e^{-\tau} n_\mathrm{e} \; v_\mathrm{p}  \;,
\end{equation}
where $\rm{T_\mathrm{CMB}}=2.725\,K$ is the average CMB temperature, $\sigma_\mathrm{T}$ is the Thomson cross-section, $n_\mathrm{e}$ is the electron density, and $v_\mathrm{p}$ is the peculiar velocity along the line of sight. In full generality, the kSZ signal includes contributions from variations in the line-of-sight velocity of electrons within the halo. However, following common practice in observational analyses~\citep[e.g.,][]{Amodeo2021_sz, Schaan2021_sz}, we assume the gas moves coherently with the halo’s bulk velocity (see Ref.~\cite{Hadzhiyska_pecvel_2023}), allowing the peculiar velocity $v_\mathrm{p}$ to be factored out of the line-of-sight integral. This simplification is commonly used in stacking analyses, where individual velocities are not precisely known. Instead, the expected signal is modeled using the root-mean-square (RMS) radial velocity of the halo sample, which we fix to $v_\mathrm{r}=1.06 \times10^{-3}c$ following Ref.~\cite{Amodeo2021_sz}.
\\
The optical depth $\tau$ along the line of sight is defined as:
\begin{equation}
\tau(\theta) =  \sigma_\mathrm{T} \int_{\rm los} n_{\mathrm{e}}\left(\sqrt{l^2\:+\:d_A(z)^2\theta^2}\right)\mathrm{d}l\;,
\end{equation}
where $\theta = |\bm{\theta}|$. For our redshift range of interest ($0.4<z<0.7$, see~\ref{subsec: ACT data}), the mean optical depth is below $0.01$ (see, Ref.~\cite{PlanckCollab_tau_2015}) and therefore we approximate $e^{-\tau}\approx1$ in Eq. \eqref{eq: kSZ}. We can then further simplify Eq. \eqref{eq: kSZ} as:
\begin{equation}
    \frac{\Delta T_{\mathrm{kSZ}}}{T_{\mathrm{CMB}}}=\tau_{\mathrm{gal}}\left(\frac{v_\mathrm{r}}{c}\right).
\end{equation}
The electron number density $n_\mathrm{e}$ in Eq. \eqref{eq: kSZ} can be derived from:
\begin{equation} 
n_\mathrm{e} = \frac{X_\mathrm{H} + 1}{2 m_\mathrm{amu}} \rho_\mathrm{gas}, \end{equation}
where $X_\mathrm{H}=0.76$ is the hydrogen mass fraction and $m_\mathrm{amu}$ is the atomic mass unit. The gas density $\rho_\mathrm{gas}$ is obtained directly from the baryonification model (see Sec.~\ref{subsec: baryonification}).\\

The kSZ signal used in this work is based on the measurements presented in Ref.~\cite{Schaan2021_sz}, which are obtained by stacking ACT DR5 and Planck CMB temperature maps at 98 GHz (f90) and 150 GHz (f150) on the positions of spectroscopic BOSS CMASS galaxies, using their reconstructed velocities. These measurements result from an observational pipeline that includes beam convolution and filtering techniques designed to isolate the kSZ signal (see Section~\ref{subsec: ACT data} for details). To model the observed profile, we apply a convolution of Eq. \eqref{eq: kSZ} with the effective beam profiles at f90 and f150. In line with Ref.~\cite{Schaan2021_sz}, we approximate each beam as a Gaussian, using a full width at half maximum (FWHM) of 2.1 arc minutes for f90 and 1.3 arcminutes for f150. Additionally, to mitigate noise from large-scale CMB variations, a compensated aperture photometry filter was applied to the observations. We reproduce the filtering applied in the data analysis by implementing the same aperture photometry filter used in Ref.~\cite{Schaan2021_sz}. This filter is defined by a weighting function \( W_{\theta_{\rm d}}(\theta) \), which takes the form:
\[
W_{\theta_{\rm d}}(\theta) =
\begin{cases}
+1 & \text{if } \theta < \theta_{\rm d} \\
-1 & \text{if } \theta_{\rm d} \leq \theta < \sqrt{2}\,\theta_{\rm d} \\
0 & \text{otherwise}
\end{cases}
\]
where \( \theta_{\rm d} \) is the aperture radius. This compensated filter is used to suppress large-scale CMB fluctuations.

We then compute the filtered kSZ temperature using~\cite{Schaan2021_sz}:
\begin{equation}
    \mathcal{T}(\theta_{\rm d})=\int \mathrm{d}^2{\bm{\theta}}\,\delta T({\bm{\theta}})\,W_{\theta_{\rm d}}({\bm{\theta}})\: ,
\end{equation}
which effectively measures the differential temperature between the inner disk and its surrounding annulus. The result is typically reported in units of \( \mu \mathrm{K}\cdot \mathrm{arcmin}^2 \), reflecting the integration over solid angle and the convention adopted in the observational analysis.

\subsection{Two-halo term}\label{subsec: 2-halo term}
To model gas profiles out to large radii, it is essential to include the contribution from the correlated large-scale structure surrounding the halo, commonly referred to as the two-halo term. This component is combined with the gravitationally bound matter inside haloes (the one-halo term) and was introduced in Eqs.~\eqref{eq: rho1h+rho2h}. While the two-halo term is subdominant in the inner regions, its contribution becomes increasingly relevant at larger radii. This is particularly important for modeling the kSZ signal, which, unlike X-ray emission, is sensitive to the diffuse gas extending well beyond the virial radius. As demonstrated in Refs.~\cite{Covone2014_2h,Hill_2h,Garcia2021_hexcl}, the contribution from the two-halo term becomes significant at distances around 2-3 times the virial radius $r_{200}$. In this regime, the kSZ effect is still measurable and thus needs to be included in our modeling pipeline. In general, the total radial three-dimensional profile for a halo of given mass $M$ can be expressed as:
\begin{equation}
    \zeta_\mathrm{tot}(r|m) = \zeta_{1\mathrm{h}}(r|m) + \zeta_{2\mathrm{h}}(r|m)\;,
\end{equation}
where $\zeta_{1\mathrm{h}}$ represents the profile of interest, and $\zeta_{2\mathrm{h}}$ accounts for the contributions from neighboring halos. \\

The two-halo term in real space can be obtained by Fourier transforming the two-halo contribution to the power spectrum:
\begin{equation}\label{eq: zeta_2h}
    \zeta_{\mathrm{2h}}(r|m)=\frac{1}{2\pi^2} \int_0^\infty \mathrm{d}k \, k^2 j_0(kr) P_{\mathrm{2h}}(k)\;,
\end{equation}
where $j_0$ is the spherical Bessel function of the first kind, and $P_{\rm 2h}(k)$ represents the two-halo contribution to the halo profile power spectrum. This spectrum is calculated using the halo model, in which the Fourier transform of the one-halo profile is given by:
\begin{equation}
    u(k,m)= 4\pi\int_0^\infty \mathrm{d}r \, r^2 j_0(kr) \zeta_\mathrm{1h}(r|m)\;,
\end{equation}
where we have simplified the 3-dimensional integral by assuming radial symmetry in the profiles used. Given $u(k,m)$, the two-halo contribution to the power spectrum is given by:
\begin{equation}\label{eq: P_2h}
    P_\mathrm{2h}(k) = b(m)P_\mathrm{lin}(k)\int_0^\infty \mathrm{d}m' \frac{\mathrm{d}n}{\mathrm{d}m'} b(m')u(k,m')\;.
\end{equation}
This expression describes the cross-power spectrum of a halo of a given mass $M$ with all other halos in the universe. Thus, $m$ denotes the mass of the halo of interest, and the integral is performed over the masses $m'$ of the neighboring halos.  Here, $P_\mathrm{lin}$ represents the linear matter power spectrum, $b(m)$ is the linear halo bias, and $\sfrac{\mathrm{d}n}{\mathrm{d}m'}$ is the halo mass function. We adopt the Sheth-Tormen bias model~\cite{Sheth_bias_1999} and the Tinker08 halo mass function~\cite{Tinker_hmf_2008}, as implemented in the Core Cosmology Library (\texttt{CCL}\footnote{\url{https://github.com/LSSTDESC/CCL}}~\cite{chisari2019_core}). We choose to integrate over the mass range $10^{10}-10^{15} M_{\odot}$, as we find that extending the range further changes the result by less than 2\%. Finally, substituting $P_{\rm 2h}(k)$ into Eq. \eqref{eq: zeta_2h}, we obtain the two-halo contribution in real space.

To reduce computational cost, we approximate the two-halo term by taking the limit $k\longrightarrow 0$ in Eq. \eqref{eq: zeta_2h}, effectively assuming that the two-halo term is independent of the detailed shape of the profile $\zeta_{\mathrm{1h}}(r|M)$. The two-halo term reduces then to:
\begin{equation}\label{eq: approx. 2-halo term}
    \zeta_\mathrm{2h}(r|m) = \overline{\zeta}_\mathrm{2h}\:(1 + b(m)\cdot \zeta(r|m))\;,
\end{equation}
where $\overline{\zeta}_\mathrm{2h}$ represents the mean background value of the profile we are considering. For example, when modeling total gas density, we use $\overline{\zeta}_\mathrm{2h} = f_{\mathrm{gas}}\times \rho_{\mathrm{m,crit}}$, with $f_\mathrm{gas}$ being the gas fraction from the model. However, for observables like the kSZ effect, where we subtract $T_{\mathrm{CMB}}$ from the profile and thus trace fluctuations rather than absolute densities, this term drops out. This expression provides an approximate form of the two-halo term, which simplifies the full derivation by assuming that the large-scale structure can be described as a uniform background modulated by the linear halo bias $b(m)$. 

To prevent unphysical contributions inside the virial radius due to overlapping halo regions, several studies introduce a halo-exclusion term that suppresses the two-halo component at small radii (see, e.g.,~\cite{Garcia2021_hexcl}). We follow this approach and define an exponential cutoff function:
\begin{equation}
    \Theta_\mathrm{e}(r|m) = 1 - \exp\left(-A_\mathrm{e} \frac{r}{r_\mathrm{vir}(m)}\right)\;,
\end{equation}
with $A_\mathrm{e}=0.4$ (see Paper I), which smoothly transitions to zero inside the virial radius. Applying this correction leads to the modified expression:
\begin{equation}
    \zeta_\mathrm{2h}(r|m) = \Theta_\mathrm{e}(r|m) \cdot \bar{\zeta}_\mathrm{2h} \cdot \left(1 + b(m) \cdot \zeta_\mathrm{lin}(r) \right)\;.
\end{equation}
The approximate two-halo term serves as a computationally efficient alternative to the exact derivation, which involves integrating over the power spectrum. We note that the approximation captures the large-scale behavior of the full result to within 1–2\% in the total profile. This level of precision is sufficient for our modeling purposes, particularly given the statistical uncertainties in current observational data.

We adopt this approximation for the two-halo term and demonstrate in Appendix~\ref{appendix: two-halo} that it provides a sufficiently accurate description. We also evaluate the effect of including the halo-exclusion correction and compare it to the case without the halo-exclusion and show that the effect is minimal.

\subsection{Gas fractions}\label{subsec: gas fractions}
Another observable that has been shown to be essential for testing cosmological models and feedback processes in structure formation (see Refs.~\cite{Ettori_fgasCosmoProbe_2009, McCarthy_fgasBAHAMAS_2017}) are gas fractions. In this subsection, we briefly describe gas fractions and how we model them with BFC.

The gas fraction, $f_{\rm gas}$, is commonly defined as the ratio of the gas mass to the total mass within a specific radius, typically $R_{500}$ or $R_{200}$ :
\begin{equation}
f_\mathrm{gas,500} = \frac{M_{\rm gas}(< R_{500})}{M_{\rm tot}(< R_{500})} = \frac{\int_0^{R_{500}}  \mathrm{d}r'\rho_\mathrm{gas}(r') r'^2}{\int_0^{R_{500}} \mathrm{d}r' \rho_\mathrm{tot}(r') r'^2 }\;,
\end{equation}
where \( M_\mathrm{gas} \) and \( M_\mathrm{tot} \) represent the gas mass and total mass enclosed within a radius \( r \), and \( \rho_\mathrm{gas}(r) \) and \( \rho_\mathrm{tot}(r) \) are the gas density and total density profiles, respectively. The total density \( \rho_\mathrm{tot}(r) \) is computed from the sum of all individual components as:
\begin{equation}
\rho_\mathrm{tot}(r) = \rho_\mathrm{gas}(r) + \rho_\mathrm{cga}(r) + \rho_\mathrm{sga}(r) + \rho_\mathrm{dm}(r) + \rho_\mathrm{2h}(r)\;,
\end{equation}
where $\rho_\mathrm{gas}$, $\rho_\mathrm{cga}$, $\rho_\mathrm{sga}$, $\rho_\mathrm{dm}$, and $\rho_\mathrm{2h}$ represent the density profiles of gas, central galaxies, satellite galaxies, dark matter, and the two-halo term, respectively. We compute all individual profiles within the baryonification model, as discussed in detail in Section~\ref{subsec: baryonification}.

Feedback mechanisms, such as AGN and stellar feedback, impact gas fractions by redistributing baryons. These processes are included in BFC through modifications to the gas density (and subsequent back-reaction effects on the dark matter component) which depend on halo mass and redshift. Observationally, $f_{\rm gas,500}$ decreases at low masses due to strong feedback, while at high masses, it approaches but remains below the cosmic baryon fraction, $f_\mathrm{b} = \Omega_\mathrm{b} / \Omega_\mathrm{m}$, because a fraction of the baryons resides in stars.

\section{Hydrodynamical Simulations}\label{sec: hydro sims}
In this work, we use hydrodynamical simulations primarily to provide context and qualitative comparisons for our results. Specifically, we focus on the FLAMINGO suite of simulations, which we briefly describe below.
\\

\noindent\textbf{FLAMINGO Simulations:} The FLAMINGO suite (Ref.~\cite{Schaye2023_flamingo}) consists of 21 cosmological hydrodynamical simulations, covering two main box sizes (1 Gpc and 2.8 Gpc) and three particle mass resolutions ($1.34 \times 10^{8}$, $1.07 \times 10^{9}$, and $8.56 \times 10^{9}$ $\mathrm{M}_\odot$). The simulations explore nine distinct astrophysical feedback models and eight different cosmologies to robustly test baryonic effects on large-scale structures. In this work, we use two simulations from the FLAMINGO suite: the high-resolution fiducial run ($m8$) and the simulation with the strongest feedback ($f_\mathrm{gas}-8\sigma$).\\
The high-resolution simulation features $N=3600^3$ dark matter and baryonic particles, with an average gas particle mass of $m_\mathrm{gas}=1.34\times 10^8\,\mathrm{M}_\odot$, a dark matter particle mass of $m_\mathrm{CDM}=7.06\times 10^8\,\mathrm{M}_\odot$, and a comoving gravitational softening length of $\varepsilon_{\mathrm{com}}=11.2\mathrm{ckpc}$. The $f_\mathrm{gas}-8\sigma$ simulation employs a lower resolution of $N=1800^3$ particles, with $m_\mathrm{gas}=1.07\times 10^9\,\mathrm{M}_\odot$, $m_\mathrm{CDM}=5.65\times 10^9\,\mathrm{M}_\odot$, and $\varepsilon_{\mathrm{com}}=22.3\mathrm{ckpc}$. Both simulations are run in a comoving box of side length $L=1\,\rm Gpc$.\\
Both simulations incorporate advanced subgrid models for baryonic physics, including stellar and AGN feedback. The $m8$ model is calibrated to match observed gas mass fractions in low-redshift clusters and galaxy stellar mass functions at $z=0.0$ (see Ref.~\cite{Kugel_FlamingoCalib_2023}), while the simulation with the strongest feedback is calibrated to the same low-redshift clusters but with $8\sigma$ smaller gas-fractions.\\

\section{Observational data}\label{sec: Obs data}
In this section, we describe the observational datasets used to constrain and test our model. We first compare predictions against two-dimensional measurements of the kinematic Sunyaev-Zel'dovich (kSZ) effect and gas fractions derived from X-ray observations, before turning to three-dimensional gas density and pressure profiles from massive galaxy clusters. The employed datasets are summarized below.

\subsection{ACT kSZ measurements}\label{subsec: ACT data}

To provide direct constraints on the hot gas distribution in halos, we incorporate measurements of the kinematic Sunyaev-Zel’dovich (kSZ) effect from Ref.~\cite{Schaan2021_sz}, which used data from the Atacama Cosmology Telescope (ACT) at 98 GHz and 150 GHz in combination with galaxy catalogs from the Baryon Oscillation Spectroscopic Survey (BOSS). These measurements probe the line-of-sight momentum of ionized gas, providing direct sensitivity to the gas density distribution when combined with reconstructed galaxy velocities.

The analysis targets halos spanning a broad mass range from $M_{200} \sim 10^{13}$ to $10^{14} \, \mathrm{M}_\odot$ (with mean mass $\overline{M}_{200} = 3 \times 10^{13} \, \mathrm{M}_\odot$) and redshifts in the range $z \sim 0.4$–$0.6$ (mean redshift $\overline{z} = 0.55$).

To isolate the weak kSZ signal from other astrophysical contributions, a matched-filtering technique is applied to the ACT maps. This approach suppresses large-scale fluctuations from the primary CMB and other foregrounds, enhancing sensitivity to the small-scale kSZ signal. The filtered maps are then stacked at the positions of BOSS galaxies.

To reconstruct galaxy line-of-sight velocities, the study uses a Wiener-filtered velocity field derived from the BOSS galaxy density field, based on linear perturbation theory and the continuity equation. This method assumes a $\Lambda$CDM cosmology and provides unbiased estimates of large-scale velocity modes, which are essential for recovering the kSZ signal.

The stacking procedure aligns individual kSZ measurements using the reconstructed velocities, allowing the extraction of a statistically significant radial kSZ profile. This profile is obtained by averaging the stacked signals in concentric annuli centered in the CMASS galaxies. Systematic uncertainties, residual foreground contamination, and instrumental biases are carefully corrected to ensure a robust measurement of the kSZ-induced temperature fluctuations.

\subsection{X-ray gas fractions}\label{subsec: x-ray fgas}
To complement the kSZ measurements, we include observational constraints on the hot gas mass fraction across a broad range of halo masses. Specifically, we utilize two distinct datasets. The first dataset consists of the most recent measurements of gas fractions obtained from the eROSITA survey (Ref.~\cite{Popesso2024_fgas}). For the second dataset we follow Ref.~\cite{Giri2021_BCemu}, where gas fractions reported by Refs.~\cite{Lovisari2015_fgas, Gonzalez2013_fgas, Sanderson2013_fgas, Sun2009_fgas, Vikhlinin2006_fgas} were combined.

\subsubsection{eROSITA gas fractions}\label{subsubsec: eFEDS fgas}
Ref.~\cite{Popesso2024_fgas} examines the hot gas content in halos, from Milky Way-like galaxy groups to massive clusters, using X-ray observations from the eROSITA Final Equatorial Depth Survey (eFEDS). The analysis covers halos in the mass range $M_{500} \sim 10^{12} - 10^{15}, \mathrm{M}_\odot$ and redshift range $z \sim 0.01 - 0.2$, based on optically selected groups and clusters from the Galaxy And Mass Assembly (GAMA) survey.

To probe the typically faint X-ray signal in low-mass systems, the study employs a stacking approach: X-ray surface brightness profiles are extracted by averaging eROSITA data at the locations of many GAMA-selected groups within narrow mass bins. This enables the detection of hot gas in systems that are individually too faint to be resolved in X-rays. In contrast to samples limited to individually detected sources, typically biased toward the most X-ray-luminous systems, stacking provides a more representative view of the population by including lower-surface-brightness objects. The resulting stacked profiles are modeled to reconstruct average electron density profiles, from which the total hot gas mass is derived within characteristic radii such as $R_{500}$ and $R_{200}$. The gas fraction, $f_{\rm gas}$, is defined as the ratio of hot gas mass to total halo mass.

One key assumption in this method is that the faint systems included via stacking are correctly described by the same models used to fit brighter systems. If faint sources are systematically different, e.g., due to different thermodynamic structure or background contamination, this could bias the inferred gas fractions. However, stacking mitigates the selection bias associated with X-ray detection thresholds, which otherwise tend to exclude low-luminosity groups.

The total halo masses used in this study are derived from optical group catalogs, primarily using abundance matching or richness–mass relations calibrated against the \textit{Magneticum} simulation~\cite{Dolag_Magneticum_2016}. These indirect methods introduce systematic uncertainties, particularly at the low-mass end where the scatter in the mass–observable relation is large. Due to the larger uncertainties and potential biases in the halo mass estimates at the low-mass end, we exclude data points with $M_{500} < 10^{13} \mathrm{M}_\odot$ from our analysis.

\subsubsection{Compilation of pre-eROSITA X-ray gas fractions}\label{subsubsec: sambit fgas}
Ref.~\cite{Giri2021_BCemu} combined a large number of X-ray gas fraction observations from different data sets, as described below:
\begin{itemize}
    \item $\textbf{Lovisari+15}$~\cite{Lovisari2015_fgas} analyzed a sample of 20 galaxy clusters and groups, spanning a mass range of $3\times10^{13}\;\mathrm{M}_\odot < M_{500} < 2\times 10^{14} \;\mathrm{M}_\odot$ and a redshift range of $0.010 \leq z \leq 0.035$. Halo masses were estimated under the assumption of hydrostatic equilibrium, using density and temperature profiles from XMM-Newton X-ray observations. Gas masses were derived by integrating the electron density profiles inferred from X-ray surface brightness modeling.
    \item $\textbf{Gonzalez+13}$~\cite{Gonzalez2013_fgas} investigated 15 galaxy clusters with halo masses ranging from \(1\times10^{14} \;\mathrm{M}_\odot < M_{500} < 2\times 10^{15}\;\mathrm{M}_\odot\), and a redshift range extending up to \(z \approx 0.2\) (median \(z \approx 0.1\)). Total mass estimates were obtained via the total baryonic content method, combining stellar masses from near-infrared observations and gas masses from Chandra X-ray data.
    \item $\textbf{Sun+09}$~\cite{Sun2009_fgas} focused on a sample of 43 galaxy groups in the mass range \(1\times10^{13}\;\mathrm{M}_\odot < M_{500} < 1\times 10^{14}\;\mathrm{M}_\odot\), with redshifts between \(0.012 \leq z \leq 0.12\). Total masses were estimated assuming hydrostatic equilibrium using temperature and density profiles derived from Chandra observations. Gas masses were computed by integrating the resulting gas density profiles.
\end{itemize}

The selected gas fractions are broadly consistent with those used to calibrate the fiducial FLAMINGO simulation.
Across all samples, gas fractions are reported with statistical uncertainties on the mean, but do not explicitly account for systematic uncertainties such as the hydrostatic mass bias ($b_\mathrm{hse}$). Hydrostatic masses tend to be biased low by roughly 20-30\% (Ref.~\citep{Munoz_bhse_2024}) due to unaccounted non-thermal pressure support from turbulence or bulk motions. In contrast, weak lensing and dynamical mass estimates are largely unbiased within current uncertainties. Second, selection biases, especially in X-ray-selected samples, can affect the observed gas fractions, as the sample may preferentially include systems with brighter or more centrally concentrated X-ray emission. Gas mass estimates themselves are typically derived from X-ray observations and depend on assumptions about the gas density profile, the conversion between X-ray emission and gas density, and the gas metallicity. In addition, calibration uncertainties in X-ray instruments further contribute to the total measurement uncertainty.

\subsection{ X-COP gas density and pressure profiles}\label{subsec: xcop sample}
To test the three-dimensional structure of the intracluster medium, we compare our model predictions to the gas density and pressure profiles observed in the XMM-Newton Cluster Outskirts Project (X-COP)~\cite{Eckert_XCOPmain_2017, Ghirardini2019_rho_pth}.\\ Ref.~\cite{Ghirardini2019_rho_pth} analyzed the radial electron number density and pressure profiles of the intracluster medium (ICM) in 12 nearby galaxy clusters\footnote{Throughout this work, we define \textit{galaxy groups} as halos with $M_{200} < 10^{14}, \mathrm{M}_\odot$, and \textit{galaxy clusters} as those with $M_{200} \geq 10^{14}$.} taken from the X-COP sample. This sample consists of massive clusters with halo masses in the range $M_{200}\sim 3\times10^{14}-1.5\times10^{15}\mathrm{M}_\odot$, where the mean mass is $M_{200}\approx 9.7\times 10^{14}\mathrm{M}_\odot$ (see Ref.~\cite{Ettori_xcop_2019}), selected to be among the brightest clusters in the Planck SZ catalog~\citep{Planck_SZCat_2013}. The clusters span a redshift range of $z=0.04-0.1$ ($\overline{z}=0.065$), making them ideal for detailed thermodynamic studies out to large radii.

By combining X-ray observations from XMM-Newton with SZ data from Planck, robust radial profiles for electron number density and pressure are derived. The electron number density is inferred from X-ray surface brightness profiles, which depend on the emission measure of the hot gas. Meanwhile, the pressure profile is constrained using the thermal Sunyaev-Zel'dovich effect, which measures the comptonization of CMB photons due to the thermal motion of electrons in the ICM. These measurements allow for a comprehensive thermodynamic characterization of the ICM, extending from the central regions ($\sim0.02R_{200}$) to the outskirts at $\sim1.5\times R_{200}$. 

For the present analysis, we prioritize like-for-like modeling of \emph{individual} clusters with tight control over selection and deprojection systematics. X-COP’s well-resolved, deprojected gas-density and pressure profiles, derived uniformly for all targets, enable this. Complementary higher-$z$ samples (e.g., \cite{McDonald_xray_2014,McDonald_xray_2017}) broaden redshift and mass coverage but often trade per-cluster outskirts signal-to-noise for sample size, and in some cases do not provide individual thermodynamic profiles. We therefore leave a detailed, forward-modeling based comparison of X-ray surface-brightness profiles to future work.

\section{Likelihood and Inference}\label{sec: Inference}

\subsection{Likelihood}\label{subsec: likelihood}
To constrain the parameters of our model, we employ a Bayesian analysis framework. Specifically, we use a Gaussian likelihood to compare model predictions with observational data. We assume that the different probes are independent, enabling the total likelihood to be expressed as a product of individual likelihoods. The Gaussian log-likelihood function, in its most general form, is given by:

\begin{equation}
\log \mathscr{L} \propto -\frac{1}{2}(\bm{d} - \bm{m})^\mathrm{T} \mathbf{C}^{-1}(\bm{d} - \bm{m}),
\end{equation}
where $\bm{d}$ represents the data vector, $\bm{m}$ the corresponding model vector, and $\mathbf{C}^{-1}$ the inverse covariance matrix. This form accounts for the full covariance structure of the data, including any correlations between different data points. In the following, we describe each of the employed likelihoods in more detail.\\

\noindent\textbf{kSZ+X-Ray fit:} For the joint fit of the kSZ signal and gas fractions (for both samples used), we assume that the two observations are independent, allowing us to sum the log-likelihoods of the two probes. \\
For the kSZ signal, we follow Ref.~\cite{Schaan2021_sz} and employ the joint covariance matrix of the two frequency observations, including their cross-covariance.\\
For the gas fraction measurements, we use the error bars on the measurements as diagonal entries in the covariance matrix, neglecting any correlations between data points. \\

Since the two probes trace different redshifts ($z = 0.55$ for the kSZ signal and $z \sim 0.1$ for the gas fractions), we evaluate the model at their respective redshifts when computing each likelihood. While the underlying model parameters may in principle evolve with redshift, we assume they remain approximately constant between $z = 0$ and $z = 0.55$. This assumption is motivated by the lack of significant evolution in the best-fit parameters in Paper I~\cite{Schneider2025_code} (see Fig. 11). Importantly, this choice applies only to the gas-profile parameters themselves; all other redshift dependencies (e.g. mass scaling relations,  distances, and profiles) are fully accounted for. Pseudo-evolution of halo masses (e.g., Ref.~\cite{Diemer_pseudo_2013}) could, in principle, introduce apparent redshift trends; however, over the narrow redshift range considered here, such effects are expected to be negligible relative to the statistical uncertainties.\\

\subsection{Sampling}\label{subsec: sampling}
We sample the likelihoods using the importance nested sampler \textsc{Nautilus}\footnote{\url{https://nautilus-sampler.readthedocs.io/en/latest/}.}, which is based on the nested sampling algorithm (see Ref.~\cite{Skilling_Nested_2007}). A key advantage of nested sampling is that it naturally provides posterior samples as a byproduct of the algorithm, making it especially useful for complex, high-dimensional, and multi-modal distributions.\\
\textsc{Nautilus} improves upon standard nested sampling by dynamically allocating live points and employing an adaptive ellipsoidal decomposition to efficiently sample complex distributions. Instead of relying on fixed clusters, it refines its sampling strategy based on local posterior structure, allowing it to resolve sharp likelihood features while maintaining computational efficiency. This approach ensures that the algorithm remains flexible and adaptive, reducing unnecessary evaluations in less relevant regions of parameter space. \\
\textsc{Nautilus} depends on a number of hyperparameters that control the accuracy, sampling efficiency and the number of likelihood evaluations. For our analysis, we use the following hyperparameter settings: 2'000 live points (\textbf{nlive}), 20'000 effective samples (\textbf{neff}), and a pruning fraction of live points (\textbf{flive}) of 0.01. We validate these by ensuring that our results are unchanged for a number of different choices.
\\

We assume physically motivated uniform priors for most parameters, with the exception of the hydrostatic mass bias, for which we impose a Gaussian prior centered at 0.26 with a standard deviation of 0.07. This prior is based on the analysis of Ref.\cite{Hurier_hydmassbias_2018}, which combines CMB lensing and SZ measurements, and is consistent with other direct estimates from X-ray and lensing data (e.g., Ref.\cite{Douspis_hydmassbias_2019}). A full overview of the adopted priors is provided in Table~\ref{tab: parameters}.

We begin with the full set of model parameters and perform a reduction analysis to identify the subset most relevant for the observables considered. This selection is guided by the physical relevance of each parameter for the kSZ signal and gas fraction measurements, as well as insights from the companion paper~\cite{Schneider2025_code}. In the final reduced model, we retain the parameters most sensitive to the hot gas distribution: $M_{\rm c}$, $\mu$, and $\delta$ (governing the gas profile shape), along with $\overline{M}_{200,\rm ksz}$ (characterizing the mean halo mass of the kSZ sample). The hydrostatic mass bias parameter $b_{\rm hse}$ is additionally included when fitting to pre-eROSITA gas fraction data. A detailed discussion of the parameter reduction and the impact of fixing other parameters is provided in Appendix~\ref{appendix: param reduc}.

\section{Results}\label{sec: results} In this section, we present the results of fitting our baryonification model to observational data and we derive constraints on baryonic feedback.\ Section~\ref{subsec: 2d profiles} presents the joint analysis of the stacked kSZ signal and X-ray gas fractions, focusing on the consistency across different observables. Section~\ref{subsec: impl. baryonic feedback} presents the implications of our findings for baryonic feedback, including the predicted three-dimensional gas and pressure profiles and the matter power spectrum suppression resulting from our best-fit model.

\subsection{2D profiles}\label{subsec: 2d profiles} We begin by jointly fitting our model to the stacked radial kSZ measurements and X-ray gas fraction data. We consider two distinct gas fraction datasets: first, the recent eROSITA-based measurements (see Sec.~\ref{subsubsec: eFEDS fgas}), and second, the pre-eROSITA compilation described in Sec.~\ref{subsubsec: sambit fgas}.

\subsubsection{kSZ + eROSITA gas fractions}\label{subsubsec: ksz + fgas}

\begin{figure*}
  \begin{center}
    \includegraphics[width=\textwidth]{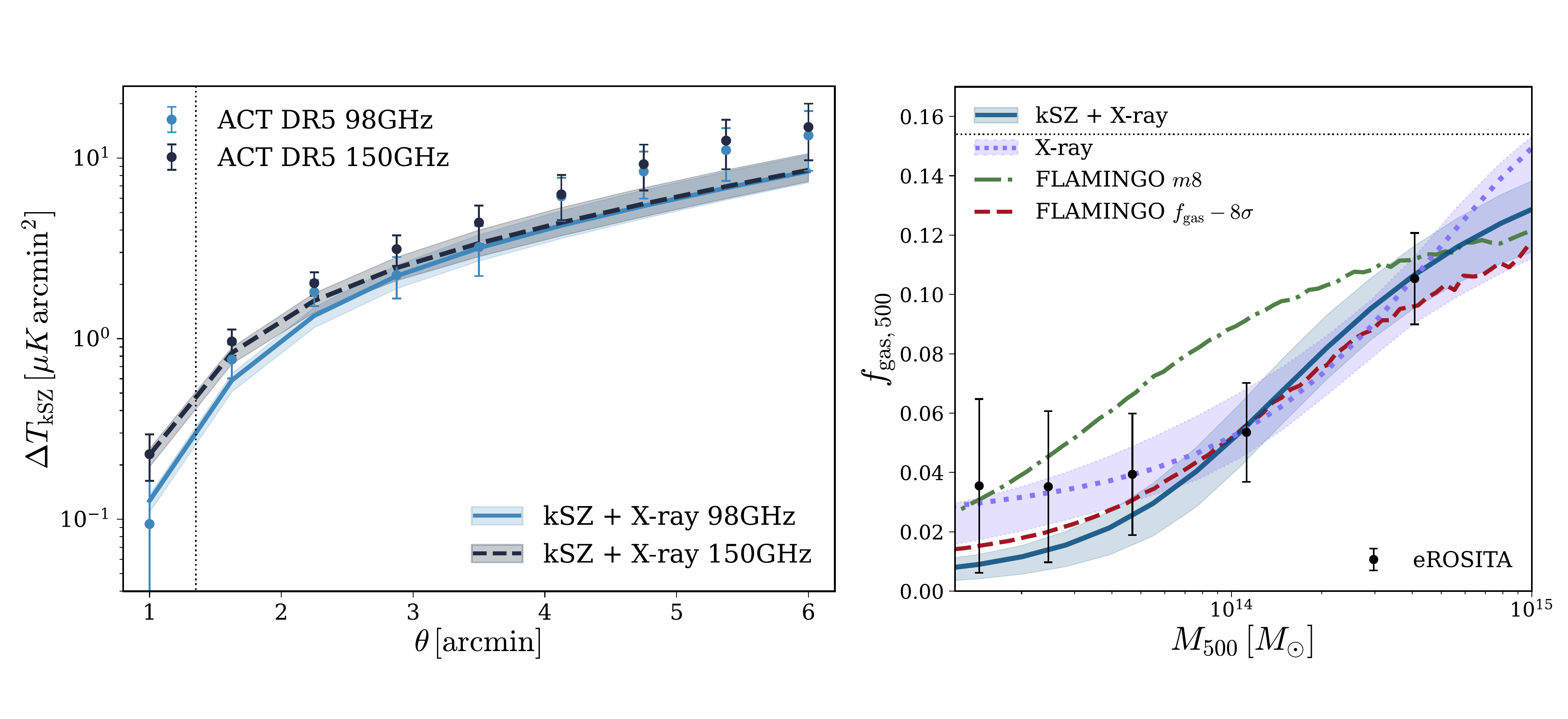}\\
    \caption{\textbf{Joint fit to the kSZ signal and eROSITA gas fractions.} \textit{Left:} Measured kSZ signal from ACT at 98 GHz (light-blue) and 150 GHz (dark-blue), compared to the BFC model prediction (light-blue solid and dark-blue dashed lines, respectively) based on the best-fit parameters from the joint kSZ + eROSITA fit. The shaded regions indicate the 68\% confidence interval. \\ \textit{Right:} Predicted gas fractions at $r_{500}$ as a function of halo mass from the joint fit (solid blue line with 68\% confidence band), compared to eROSITA measurements (black points). The dotted purple line shows the fit to gas fractions alone. For comparison, predictions from two FLAMINGO simulation models are included: the fiducial feedback $m8$ model (dash-dotted green) and the strong feedback $f_{\mathrm{gas}}$–8$\sigma$ model (dashed red).} \label{fig: eksz_fit}
  \end{center}
\end{figure*}

We now present the results of the joint fit to the kSZ signal and the eROSITA gas fractions, following the procedure described in Section~\ref{subsec: likelihood}. The fit includes four free parameters in total: three describing the gas profile, $M_\mathrm{c}$, $\mu$, and $\delta$, and one for the mean halo mass of the kSZ sample, $\overline{M}_{200,\mathrm{ksz}}$. This joint analysis builds on an initial model with nine\footnote{Since $\alpha_{\mathrm{nt}}$ is only needed for the pressure profiles, we do not include it here.} free parameters. To reduce complexity and improve robustness, we performed a parameter reduction analysis and found that the data can be well described using this reduced set. Full details are given in Appendix~\ref{appendix: param reduc observations}.

Even when leaving the halo mass $\overline{M}_{200,\mathrm{ksz}}$ free in the fit, we recover a value fully consistent with the CMASS mean mass. The best-fit value agrees well with the independent estimate from Ref.~\cite{McCarthy_kSZ2024}, who find $\log{10}(M_{500c}/\mathrm{M}_\odot) = 13.34 \pm 0.04$, corresponding to $\log_{10}(M_{200c}/\mathrm{M}_\odot) \approx 13.52$ under the assumption of an NFW profile. This agreement is further supported by the posterior distribution shown in Appendix~\ref{appendix: posterior comparison} (see Figure~\ref{fig: corner 3p eksz comparison}).

Figure~\ref{fig: eksz_fit} shows the outcome of the joint fit, with the left-hand showing the kSZ signal and the right-hand displaying the predicted gas fractions $f_\mathrm{gas,500}$ evaluated at $r_{500}$. As discussed above, we exclude the three lowest gas fraction data points due to uncertainties in the estimated halo masses, but our results are insensitive to this choice.

Focusing first on the kSZ signal, the solid and dashed lines in the left-hand-side panel represent the model predictions at 98 GHz and 150 GHz, respectively, with shaded regions corresponding to the 68\% confidence interval. As can be seen, the BFC prediction provides a good fit to the ACT data at both frequencies, particularly in the central regions. At large radii, the model systematically underpredicts the data, although, as discussed below, this has no impact on the goodness-of-fit since these outer data points are strongly correlated (see Figure 7 in Ref.~\cite{Schaan2021_sz}). Furthermore, in the outer regions, the kSZ signal is largely dominated by the two-halo term, which depends primarily on the halo mass and thus exhibits only an indirect sensitivity to the gas profile parameters (see Eq.~\eqref{eq: approx. 2-halo term}).

We note that in our BFC predictions we assume that all CMASS galaxies are centrals and neglect the contribution from satellite galaxies. In Ref.~\cite{McCarthy_kSZ2024}, it was shown that including satellites increases the predicted kSZ signal by up to 20–30\% (see Figure 4 in Ref.~\cite{McCarthy_kSZ2024}). Since our model accounts only for central galaxies, we do not directly compare our predicted kSZ signal to the FLAMINGO results (Ref.~\cite{McCarthy_kSZ2024}), which include both centrals and satellites. A comparison between our model and the FLAMINGO kSZ measurements, highlighting the effect of satellite contributions, is provided in Appendix~\ref{appendix: satellite discussion}. As discussed there, our results remain robust to this simplification given the current statistical uncertainties, though satellite effects will need to be modeled more carefully in future high-precision analyses.

The corresponding constraints on the gas fraction from the joint fit are shown as the solid blue line with 68\% confidence intervals in the right-hand panel. While the kSZ-only fit already provides tight constraints on the $\delta$ parameter, incorporating the X-ray gas fractions additionally constrains the mass-dependent parameters $M_\mathrm{c}$ and $\mu$. The joint fit closely traces the gas fraction data and remains within $1\sigma$ of all measurements, though it tends to lie on the lower edge at low masses.

Crucially, the inclusion of the kSZ data shifts the gas fraction fit downward, particularly in the mass range around $\overline{M}_{200,\mathrm{ksz}}\approx 3\times 10^{13}\mathrm{M}_\odot$ ($M_{500}=2.1\times 10^{13}\mathrm{M}_\odot$). This suppression is clearly visible when comparing the joint fit to the gas-fraction-only fit (dotted purple line). In addition, the kSZ data recover the characteristic smooth step function in the gas fraction–mass relation, which is not apparent in the X-ray-only fit. This feature reflects the mass-dependent efficiency of AGN feedback: flattening at low masses due to strong gas removal and flattening again at high masses where the deeper potential wells retain more gas.

Despite this shift, the two fits remain statistically consistent (see the joint posterior distribution in Appendix~\ref{appendix: posterior comparison} in Figure~\ref{fig: corner 3p eksz comparison}).

The quality of the joint fit is quantified via $\chi^2$ values computed from Monte Carlo realizations of the data. For the joint kSZ and eROSITA gas fraction fit, we find $\chi^2 = 24.57$ and a PTE (Probability to exceed) of 0.41 (see Table~\ref{tab:chi-squared_statistics}), indicating a good fit and no significant tension between the two datasets.

Finally, we compare our joint fit predictions to results from hydrodynamical simulations. The $m8$ FLAMINGO simulation (dash-dotted green line), which~\cite{Schaye2023_flamingo} show reproduces pre-eROSITA data, overestimates the gas fraction at nearly all masses, with agreement emerging only at the high-mass end ($M_{500}>5\times10^{14}\mathrm{M}_\odot$). The $f_\mathrm{gas}$–8$\sigma$ simulation lies within our 68\% interval across most of the mass range, although it slightly undershoots the high-mass end and overshoots at the low-mass end.

In summary, the kSZ and eROSITA gas fraction data can be jointly explained by a consistent set of gas profile parameters. The combination of these two observables constrains both the amplitude and shape of the gas distribution, with the kSZ data playing a crucial role in limiting the gas content at low masses.

\begin{table}[ht]
    \centering
    \renewcommand{\arraystretch}{1.2}
    \begin{tabular}{lccc|ccc} 
        \hline \hline
        & \multicolumn{3}{c|}{\textbf{eROSITA}} & \multicolumn{3}{c}{\textbf{pre-eROSITA}} \\
        \textbf{Statistic} 
        & kSZ & $f_{\mathrm{gas,500}}$ & \textbf{kSZ + $f_{\mathrm{gas,500}}$} 
        & kSZ & $f_{\mathrm{gas,500}}$ & \textbf{kSZ + $f_{\mathrm{gas,500}}$} \\ 
        \hline
        $\chi^2_{\mathrm{actual}}$ 
        & 22.56 & 2.00 & \textbf{24.57} 
        & 24.54 & 2.48 & \textbf{27.03} \\
        PTE 
        & 0.22 & 0.82 & \textbf{0.41} 
        & 0.14 & 0.70 & \textbf{0.22} \\
        \hline \hline
    \end{tabular}
    \caption{%
        $\chi^2$ values and p-values (PTE) from the joint fit of the kSZ signal and X-ray gas fractions.
        The \textit{kSZ} and \textit{X-ray} columns show the $\chi^2$ and PTE for each dataset, 
        evaluated at the best-fit parameter values from the joint analysis (see Figures~\ref{fig: eksz_fit} and~\ref{fig: fgas_joint_fit_sksz}).
        The \textbf{joint fit} columns display the total $\chi^2$ and PTE from the combined analysis. 
        Results are shown for both the eROSITA (left) and pre-eROSITA (right) gas fraction datasets.
    }
    \label{tab:chi-squared_statistics}
\end{table}

\subsubsection{Compilation of pre-eROSITA X-ray gas fractions}\label{subsubsec: res pre-eROSITA results}

\begin{figure}[htbp] \centering \includegraphics[width=\linewidth]{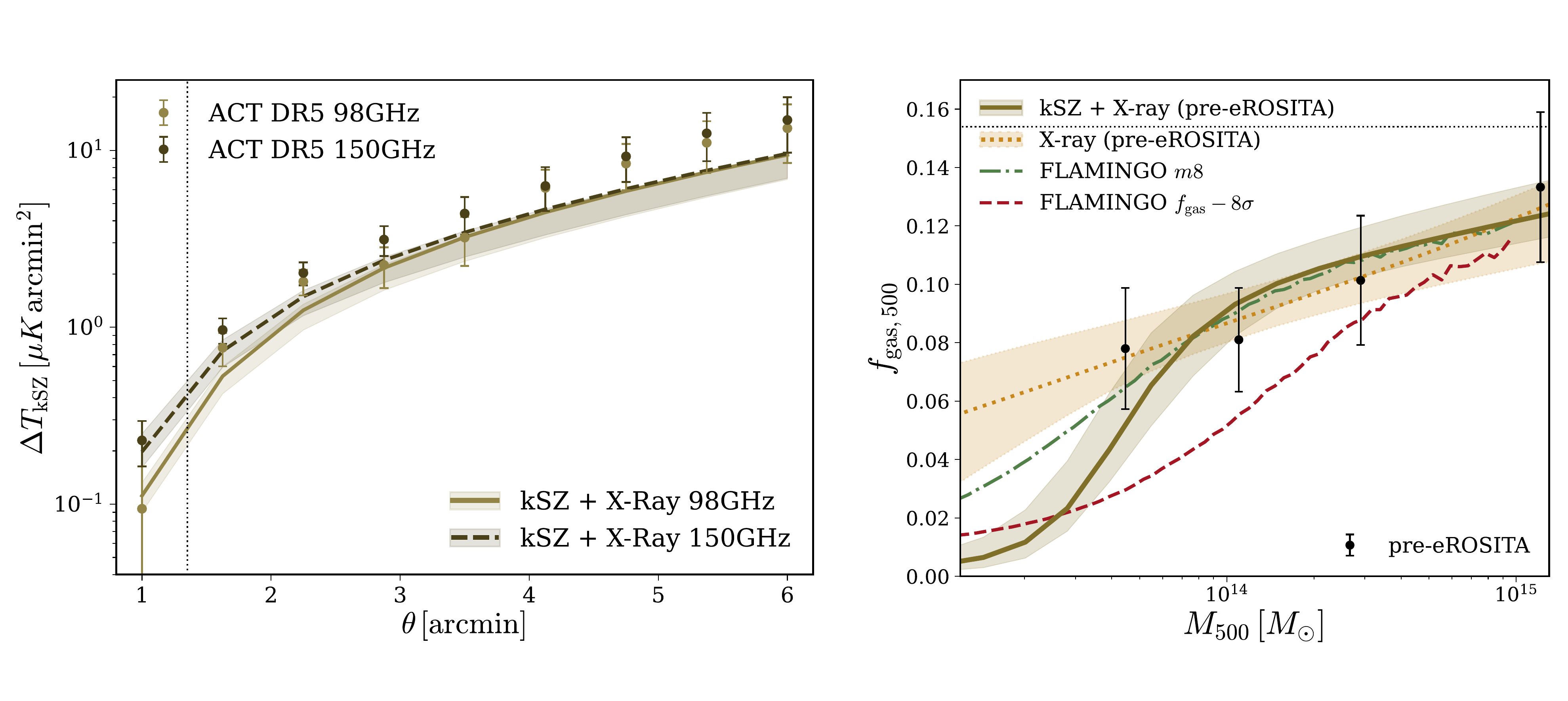} \caption{\textbf{Joint fit to the kSZ signal and pre-eROSITA gas fractions.} Same as Figure~\ref{fig: eksz_fit}, but using the pre-eROSITA X-ray gas fraction dataset.} \label{fig: fgas_joint_fit_sksz} \end{figure}

As in the previous section, we perform a joint fit of the kSZ signal with the X-ray gas fraction measurements, this time using the pre-eROSITA dataset. While the model and fitting procedure remain unchanged, we apply the hydrostatic mass bias correction to the pre-eROSITA gas fraction data. The results are presented in Figure~\ref{fig: fgas_joint_fit_sksz}, where the left-hand panel shows the fit to the kSZ signal and the right-hand panel displays the gas fractions. The solid blue line indicates the posterior mean of the joint fit, and the shaded region shows the corresponding 68\% confidence interval. Notably, we observe visible differences in the predicted kSZ signal compared to the kSZ+eROSITA fit (Figure~\ref{fig: eksz_fit}), although the overall agreement remains acceptable\footnote{Quantified using the PTE, which is 0.14 for the joint fit to the kSZ signal (corresponding to a $\chi^2 = 24.54$).}.

In the gas fraction panel, the joint fit closely follows the four available pre-eROSITA data points, all of which lie within the 68\% confidence region. However, a noticeable divergence emerges when extrapolating to lower masses, below the lowest data point at $M_{500} \sim 5 \times 10^{13}\,\mathrm{M}_\odot$. In this regime, the joint fit, driven by the kSZ signal, exhibits a clear downturn in the gas fraction, while the individual X-ray-only fit continues to predict comparatively high values. This behavior suggests that the kSZ data impose a preference for lower gas fractions at the group scale, creating tension with the trend implied by the pre-eROSITA gas fractions alone.

This discrepancy is consistent with earlier work: previous studies (e.g., Refs.~\cite{McCarthy_kSZ2024,Bigwood2024, Salcido_prefgasTension_2024}) have similarly reported challenges in jointly fitting the kSZ signal and earlier X-ray gas fraction measurements. The $\chi^2$ of the joint fit remains within acceptable bounds, yet the parameter space required to reconcile both probes results in rather extreme values, especially for $M_{\mathrm{c}}$ and $\mu$ (see Figure~\ref{fig: corner 3p eksz comparison} in Appendix~\ref{appendix: posterior comparison}). This reinforces the conclusion that the two datasets prefer markedly different regions of parameter space, with the X-ray data favoring low values of $M_\mathrm{c}$ and $\mu$, while the kSZ data prefer significantly higher values (see Figure~\ref{fig: corner 3p eksz comparison} in Appendix~\ref{appendix: posterior comparison}).

A likely explanation for this discrepancy lies in selection effects. The pre-eROSITA X-ray data were primarily derived from pointed observations and archival samples, which preferentially detect systems with high gas content, while differences in mass estimation may also contribute. This leads to a systematic overestimation of the average gas fraction, particularly at the group scale, compared to more representative samples. In contrast, the eROSITA dataset combines X-ray-selected halos with optically selected groups from the GAMA survey, providing a more balanced and less biased view of the halo population~\cite{Popesso_selectionBias_2024, Clerc_selectionBiases_2024, Comparat_selectionBias_2022}.

The comparison to FLAMINGO simulation results shows that the pre-eROSITA joint fit overshoots the gas fractions from the $f_{\mathrm{gas}}$–8$\sigma$ model across most mass scales, while aligning more closely with the $m8$ simulation in the mass range covered by the X-ray data, as expected, given that this simulation was specifically calibrated to match these gas fraction observations. However, at the lower masses relevant for the kSZ data ($M_{500} \sim 2\times10^{13}\,\mathrm{M}_\odot$), the gas fractions in the joint fit fall significantly below the $m8$ FLAMINGO prediction but agree well with the $f_{\mathrm{gas}}$–8$\sigma$ simulation, another sign of the strong suppression driven by the kSZ.

Finally, comparing the joint fits using either eROSITA or pre-eROSITA gas fractions reveals a notable difference in the shape of the gas fraction profiles at low masses. In both cases, the inclusion of kSZ data induces a downturn in $f_{\mathrm{gas,500}}$, reflecting the suppression of gas content required by the kSZ signal. However, the downturn is much steeper in the pre-eROSITA fit. This is because the pre-eROSITA gas fraction measurements remain high at low masses, forcing the model to compensate more strongly to match the kSZ data. In contrast, the eROSITA gas fractions already lie at lower values for group-scale halos, resulting in a smoother, more gradual suppression in the joint fit.

In summary, our results confirm that the pre-eROSITA gas fraction measurements are in tension with the kSZ signal, as a joint fit requires extreme parameter combinations and still fails to reproduce the low-mass suppression indicated by the kSZ data. However, when replacing the pre-eROSITA measurements with the new eROSITA gas fractions, this tension is resolved: the two datasets become mutually consistent and jointly favor models with strong baryonic feedback. This conclusion is further supported by the good agreement between our best-fit gas fractions and the predictions of the $f_\mathrm{gas}-8\sigma$ FLAMINGO simulation, which not only reproduces the kSZ signal, as shown in Ref.~\cite{McCarthy_kSZ2024}, but also closely matches our results.

\subsection{Implications for baryonic feedback}\label{subsec: impl. baryonic feedback}
Having established that our model allows us to self-consistently fit the kSZ measurements from ACT and gas fractions from eROSITA, we now examine the implications of our results for baryonic feedback effects. We proceed in two steps:  We first present the three-dimensional gas density profiles predicted by our model and compare them to results from hydrodynamical simulations and observations. In the second step, we compute the predicted matter power spectrum suppression on small and intermediate scales.

\subsubsection{Comparison of 3D Profiles to FLAMINGO and X-COP}\label{subsubsec: 3D profiles comparison}

After jointly fitting our gas model to the kSZ signal and gas fractions from eROSITA and pre-eROSITA, we examine the resulting gas density profiles using the posterior from the kSZ + eROSITA fit. We focus on two characteristic halo mass scales: the mean mass of the CMASS BOSS sample, $\overline{M}_{\mathrm{200}} = 3 \times 10^{13} \mathrm{M}_\odot$, relevant for the kSZ signal; and the mean mass of the X-COP sample, $\overline{M}_{\mathrm{200}} = 9.7 \times 10^{14} \mathrm{M}_\odot$, for which high-resolution X-ray data are available. To account for redshift evolution, we evaluate the gas profiles at the mean redshifts of the respective samples: $z = 0.55$ for CMASS and $z = 0.065$ for X-COP.

For the X-COP sample, masses determined assuming hydrostatic equilibrium have been shown to be consistent with weak-lensing determinations (Ref.~\cite{Ettori_xcop_2019}), and the inferred non-thermal pressure support is small ($\sim6\%$ at $R_{500}$). We therefore treat the X-COP masses as effectively unbiased for our purposes and do not apply an explicit hydrostatic-bias correction when modeling the individual clusters.

\begin{figure} \centering \includegraphics[width=0.65\linewidth]{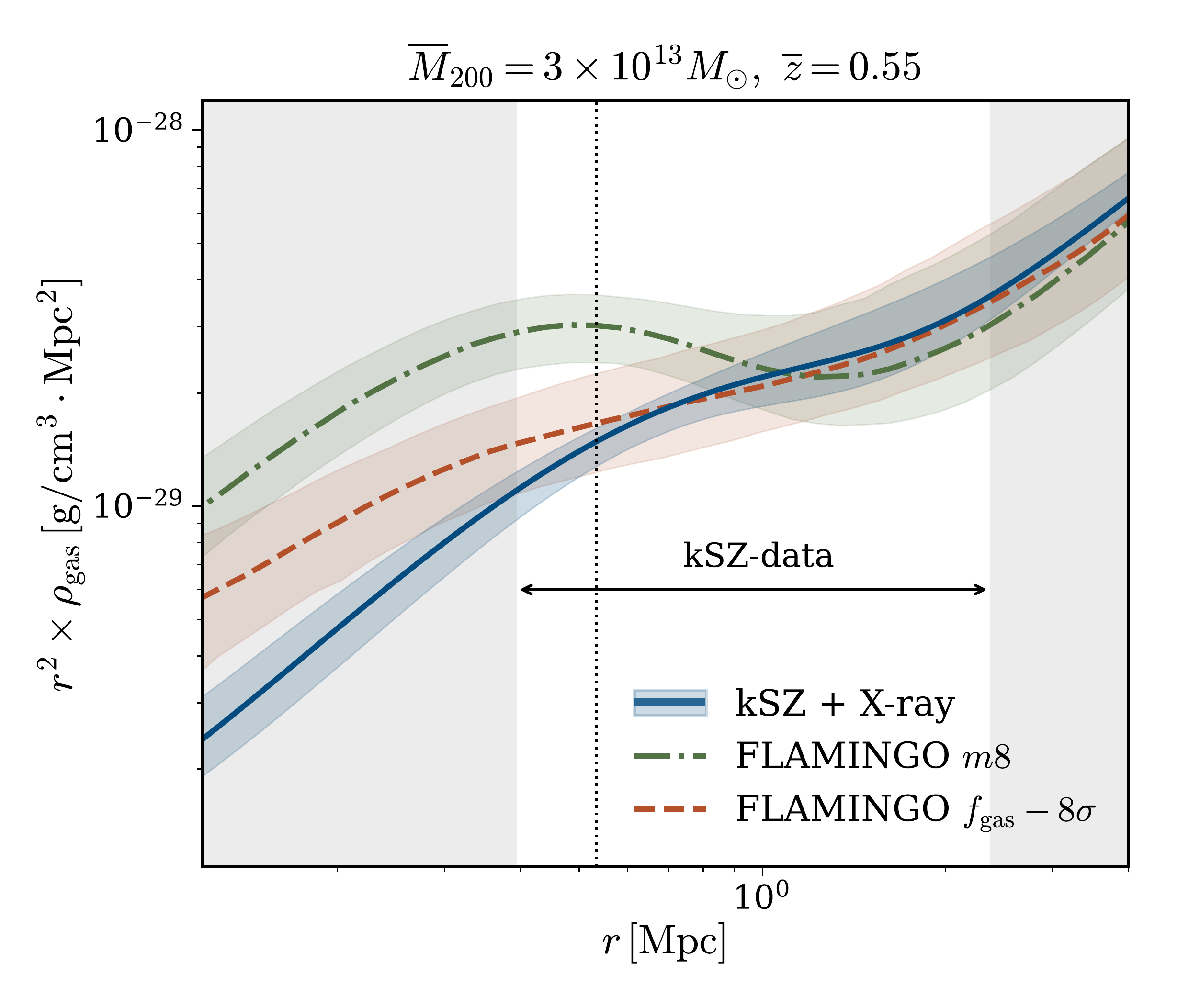} \caption{\textbf{Gas density profile at the CMASS mean mass.} Shown is the three-dimensional gas density profile at $\overline{M}_{200} = 3 \times 10^{13}\,\mathrm{M}_\odot$ and redshift $z = 0.55$, based on the posterior from the joint kSZ + eROSITA fit (solid blue line with 68\% confidence interval). The FLAMINGO $m8$ and $f_{\mathrm{gas}}$–8$\sigma$ simulation profiles are included for comparison (dash-dotted green and dashed red lines, respectively). The vertical dotted line marks the virial radius. The region directly constrained by the kSZ data is indicated, while the shaded area at smaller and larger radii denotes extrapolation beyond the observational range.} \label{fig: 3D profile low mass regime} \end{figure}

\vspace{0.5em} \noindent\textbf{Low-mass regime:} Figure~\ref{fig: 3D profile low mass regime} shows the gas density profile at the CMASS mean mass. The solid blue line denotes the posterior mean from our joint fit, with the shaded region showing the 68\% confidence interval. For comparison, we include gas density profiles from the FLAMINGO $m8$ and $f_{\mathrm{gas}}$–8$\sigma$ simulations (dash-dotted green and dashed red lines, respectively), evaluated using measured profiles from FLAMINGO halos in the mass range $\log_{10}(M_{200}/\mathrm{M}_\odot) = 13.13$–13.63, which encompasses the CMASS mean mass.\\
The radial region where the kSZ data constrains the profile is highlighted, while the shaded outer area consists of an extrapolation. Within the virial radius (indicated by the vertical dotted line), our model predicts a suppressed gas density compared to both FLAMINGO simulations, lying below even the $f_{\mathrm{gas}}$–8$\sigma$ model. At large radii ($r > 3\,\mathrm{Mpc}$), all profiles converge due to the dominance of the two-halo term, which is set by the cosmological background and thus insensitive to feedback.\\
In the intermediate region ($1\,\mathrm{Mpc} < r < 3\,\mathrm{Mpc}$), where the kSZ signal provides the strongest constraints, our model predicts gas densities that are nearly identical to those in the $f_{\mathrm{gas}}$–8$\sigma$ FLAMINGO run. This redistribution of gas from the center to the outskirts is required to reproduce the observed kSZ signal and is consistent with trends reported in recent analyses~\cite{Hadzhiyska2024, Bigwood2024}. Overall, the gas profile in this regime supports a strong feedback scenario similar to FLAMINGO $f_{\mathrm{gas}}$–8$\sigma$, with indications of a very slightly more efficient expulsion of gas from the halo center at the 1$\sigma$ level. This finding reinforces the consistency between the kSZ signal and the low gas fractions inferred from eROSITA, both of which point toward a scenario with strong baryonic feedback.\\

\begin{figure}[htbp]
\centering
\includegraphics[width=\linewidth]{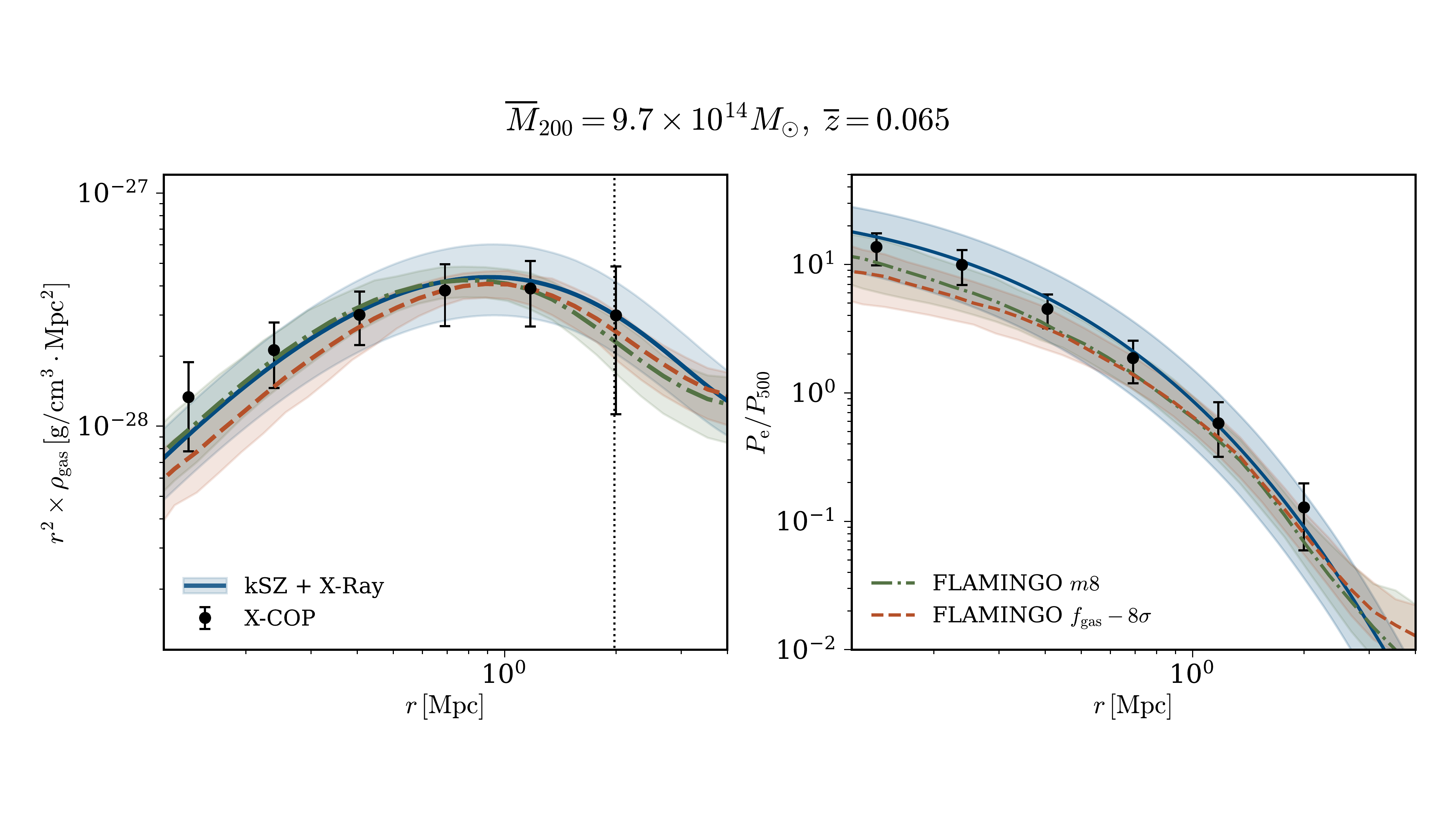}
\caption{\textbf{Gas density and pressure profiles at the X-COP mean mass.} Shown are the three-dimensional gas density (left) and normalized electron pressure (right) profiles at $\overline{M}_{200} = 9.7 \times 10^{14}\mathrm{M}_\odot$ and redshift $z = 0.065$ (mean mass and redshift from the X-COP sample), based on the posterior from the joint kSZ + eROSITA fit (solid blue lines with 68\% confidence intervals). X-COP measurements are shown as black points. The FLAMINGO $m8$ and $f_{\mathrm{gas}}$–8$\sigma$ simulation profiles are included for comparison (dash-dotted green and dashed red lines, respectively). While both simulations fall within the model uncertainty, the $m8$ run matches the gas density more closely in the cluster center, and both simulations lie slightly below the posterior mean in the pressure profile across all radii.} \label{fig: 3D profiles high mass regime}
\end{figure}

\vspace{1.5em}
\noindent\textbf{High-mass regime:} Figure~\ref{fig: 3D profiles high mass regime} shows the predicted gas density and pressure profiles for the X-COP sample. We again use the posterior from the joint kSZ + eROSITA fit and compute individual profiles for each X-COP cluster, averaging them to obtain the solid blue line. The shaded region includes both the posterior uncertainty and the scatter from the cluster-to-cluster mass variation. \\
Our model reproduces the X-COP measurements remarkably well, staying within 1$\sigma$ across the full radial range for both gas density and pressure\footnote{Normalized with $P_{500}$ following Eq. 8 in Ref.~\cite{Ghirardini2019_rho_pth}.}. This agreement is especially notable given that the model is only constrained by observables tracing the distribution of hot gas, namely the kSZ signal and X-ray gas fractions, and not by the X-COP profiles themselves. The consistency across both density and pressure strongly supports the validity of the model and demonstrates that it captures both the spatial and thermodynamic structure of the intracluster medium.\\
For comparison, we now directly use the FLAMINGO $m8$ and $f_{\mathrm{gas}}$–8$\sigma$ simulation profiles, evaluated at the mean mass of the X-COP sample at redshift $z=0.1$ ($z_2$)\footnote{We then rescaled it to the mean redshift of the X-COP sample $z=0.065$ ($z_1$) by multiplying the profiles with $(1+z_1)^3/(1+z_2)^3$}. The resulting predictions are shown as dash-dotted green and dashed red lines, respectively, with shaded bands indicating their 68\% scatter. \\
Our best-fit model aligns most closely with the FLAMINGO $m8$ simulation for the gas density in the central cluster regions, while the $f_{\mathrm{gas}}$–8$\sigma$ simulation underpredicts the core densities, suggesting that strong feedback is disfavored at these high masses. In the case of the electron pressure profile, both simulations lie within the 68\% confidence interval of our model across the full radial range. However, they systematically fall below the posterior mean at nearly all radii. Although the differences are not statistically significant given the current uncertainties, this consistent offset may point toward slightly higher electron pressure in the observed clusters than predicted by either simulation. 
Similar differences were also observed in~\cite{Braspenning_2024}, who compared pressure and density profiles from simulations to X-ray observations across a wide range of halo masses. They showed that such comparisons are highly sensitive to the assumed halo mass and to the method used to extract simulation profiles, e.g., X-ray weighting versus mass or volume weighting. As a result, observed X-ray profiles can be biased even before accounting for selection effects. This mass sensitivity implies that robust comparisons require accurately modeling the full mass distribution within the observational sample and carefully accounting for potential biases in the inferred halo masses.
As in the low-mass regime, all profiles converge at large radii where the two-halo term dominates.\\

Taken together, these results reveal a clear mass dependence in the efficiency of baryonic feedback. At low masses, our joint fit requires slightly stronger suppression of central gas than predicted by FLAMINGO’s $f_{\mathrm{gas}}$–8$\sigma$ simulation, while at intermediate and high masses the inferred profiles are broadly consistent with that strong feedback model. Only at the highest masses ($M_{500} \gtrsim 5 \times 10^{14},\mathrm{M}_\odot$) do the gas fractions and profiles approach the values of the $m8$ FLAMINGO simulation, suggesting that feedback becomes less efficient in deep gravitational potentials.

This trend is consistent with the gas fraction constraints shown in Figure~\ref{fig: eksz_fit}, which generally lie close to the $f_{\mathrm{gas}}$–8$\sigma$ simulation across most of the mass range, but begin to converge with the $m8$ simulation at the cluster scale. Notably, our model reproduces both the low-mass and high-mass gas density profiles without being explicitly fitted to the X-COP density data, only the masses of the X-COP sample are used. This consistency across mass scales highlights the robustness of the model and its ability to capture the evolving impact of baryonic feedback on halo gas content.\\

\subsubsection{Power spectrum suppression}\label{subsubsec: pk suppression}

\begin{figure}[htbp]
\centering
\includegraphics[width=0.75\linewidth]{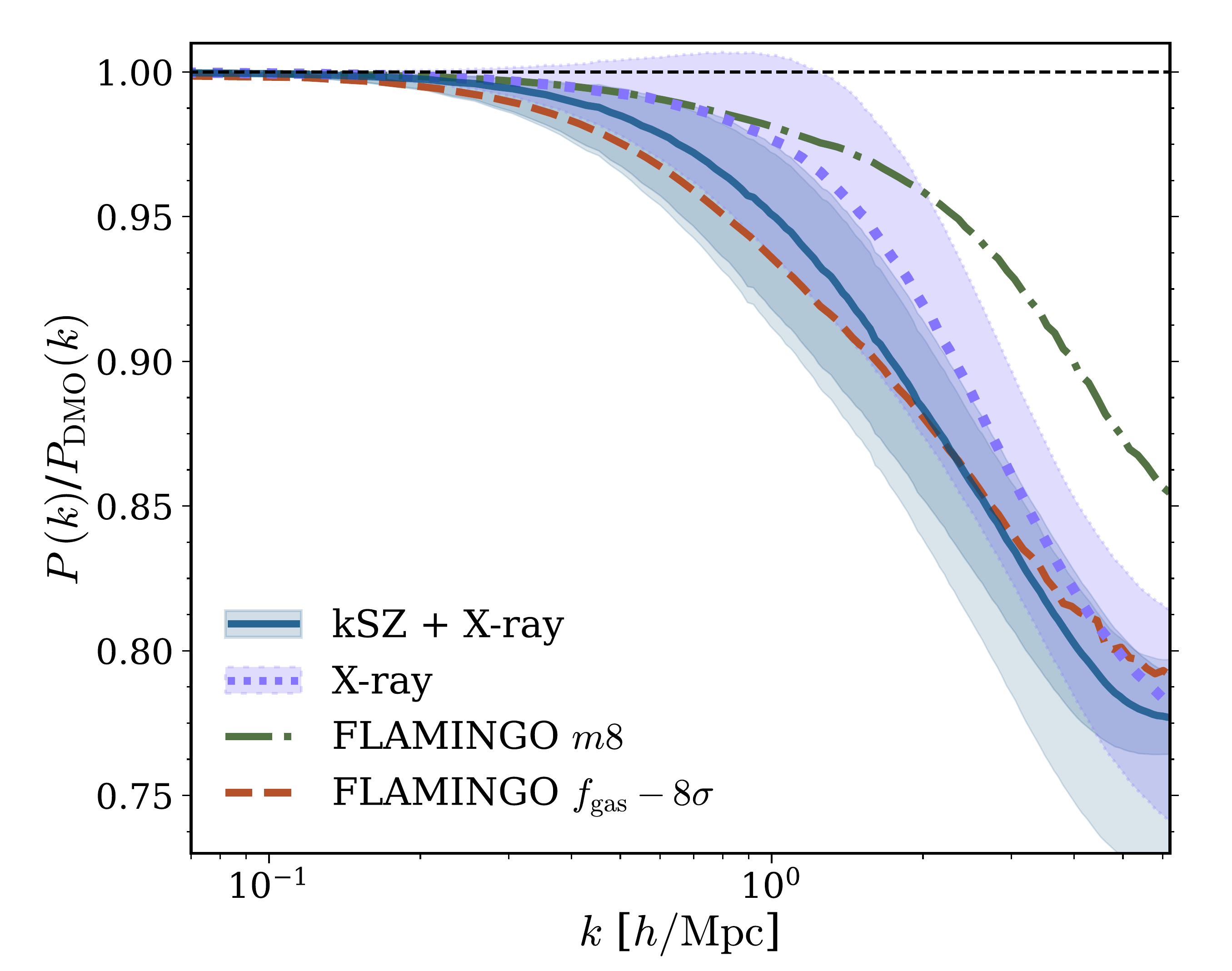}
\caption{\textbf{Matter power spectrum suppression from the joint kSZ + eROSITA fit.} Shown is the suppression of the matter power spectrum predicted using the best-fit parameters from the three-parameter joint fit to the kSZ signal and eROSITA gas fractions (solid blue line with 68\% confidence interval). The broader blue-shaded band indicates the result from a six-parameter version of the fit where stellar feedback parameters are also varied. For comparison, suppression curves from the FLAMINGO $m8$ simulation (dash-dotted green) and the  $f_{\mathrm{gas}}$–8$\sigma$ simulation (dashed red) are included, along with the suppression from a gas-fraction-only fit (purple dotted line).} \label{fig: pk suppression}
\end{figure}

To assess the impact of baryonic feedback on large-scale structure, we compute the suppression of the matter power spectrum predicted by our best-fit parameters from the joint fit to the kSZ signal and eROSITA gas fractions (see Section~\ref{subsubsec: ksz + fgas}). To this end, we use \textsc{Pkdgrav3}\footnote{\url{https://bitbucket.org/dpotter/pkdgrav3/src/master/}}~\cite{Potter_pkdgrav3_2017} to generate an N-body simulation with a box size of $L = 256\,\mathrm{Mpc}/h$ and $512^3$ particles. We then baryonify this simulation using our best-fit BFC parameters and compute the resulting matter power spectrum suppression, shown in Figure~\ref{fig: pk suppression}.

The solid blue line shows the suppression predicted by our joint three-parameter (3p) fit to the kSZ and eROSITA gas fraction data, with the shaded region indicating the 68\% confidence interval. Since the model does not constrain stellar-related parameters directly, we fix them to the best-fit values obtained from stellar mass profile fits in Ref.~\cite{Schneider2025_code} (see Appendix~\ref{appendix: param reduc observations}). To assess the impact of this assumption, we also show the broader 68\% band from a six-parameter (6p) version of the fit, where the stellar parameters $\eta$, $d\eta$, and $N_\mathrm{star}$ are allowed to vary as well. This broader band illustrates the additional uncertainty at small scales ($k \gtrsim 1\,h\,\mathrm{Mpc}^{-1}$) due to degeneracies between gas and stellar feedback.
For comparison, we include the suppression curves from the FLAMINGO $m8$ (green) and strong-feedback $f_{\mathrm{gas}}$–8$\sigma$ (red) simulations. All models agree on large scales ($k < 0.2\,h\,\mathrm{Mpc}^{-1}$), where baryonic effects are negligible. At smaller scales, our joint fit predicts significant suppression, closely tracking the FLAMINGO $f_{\mathrm{gas}}$–8$\sigma$ model and clearly deviating from the $m8$ prediction (see also Ref.~\cite{Schaller_pkFlamingo_2024} for the matter power spectrum suppression of the different FLAMINGO models)

We also show the suppression from a fit to the eROSITA gas fractions alone (dotted purple line). This result consistently predicts weaker suppression across all scales compared to the joint fit. The difference highlights the role of the kSZ data, which pushes the suppression downward, especially at $k > 0.5\,h\,\mathrm{Mpc}^{-1}$. This trend mirrors what is seen in the gas fraction–mass relation (Figure~\ref{fig: eksz_fit}), where the joint fit bends downward at low masses, again driven by the kSZ.
For completeness, we show the suppression derived from the joint fit using pre-eROSITA gas fractions in Appendix~\ref{appendix: pre-erosita pk sup}. Consistent with the higher gas fractions in the older dataset, this version predicts weaker suppression and aligns more closely with the fiducial FLAMINGO $m8$ model, as expected for a scenario with weaker baryonic feedback.

Recent weak lensing results from KiDS-Legacy suggest minimal baryonic suppression~\citep{wright_kids_2025}, but mainly probe $k \sim 1\;h\mathrm{Mpc}^{-1}$ and not the deeply non-linear regime. Its one-parameter feedback model limits flexibility at smaller scales. Unlike KiDS-1000~\citep{asgari_kids_2021}, which allowed stronger suppression, KiDS-Legacy’s preference for weaker feedback (near or below $m8$) should not be interpreted as a direct constraint on smaller-scale suppression.

Taken together, our results demonstrate that the combination of kSZ and eROSITA X-ray data provides strong and self-consistent constraints on baryonic feedback. The suppression of power predicted by our model matches the $f_{\mathrm{gas}}$–8$\sigma$ FLAMINGO simulation across a wide range of scales, while disfavoring  the $m8$ model based on multiple independent observables. Crucially, this consistency extends beyond clustering: the same best-fit parameters also reproduce the observed gas fractions and gas density profiles. Unlike previous studies that inferred strong feedback primarily from kSZ or weak lensing alone, our analysis shows that the addition of eROSITA gas fractions enables consistency with X-ray constraints as well.

\section{Conclusions}\label{sec: conclusion}

In this work, we constrain a physically motivated baryonification (BFC) model for the hot gas distribution in halos by jointly fitting ACT kinematic Sunyaev-Zel’dovich (kSZ) measurements and X-ray gas fractions from both eROSITA and pre-eROSITA samples. This analysis builds on the baryonification framework developed in Paper I~\cite{Schneider2025_code}, which allows for self-consistent modeling of Dark Matter and baryons using a reduced set of physically interpretable parameters.

We find that the combination of kSZ and eROSITA data yields self-consistent constraints on baryonic feedback strength. The joint fit simultaneously recovers the amplitude and shape of the gas fraction–mass relation and constrains the radial gas distribution, particularly at low halo masses where the kSZ signal provides unique sensitivity.

In contrast, a joint fit using pre-eROSITA gas fractions leads to significant tension with the kSZ signal, favoring parameter values that are incompatible with the suppression seen at low masses. This discrepancy may be driven by selection biases in the older X-ray datasets, which predominantly include gas-rich systems, though differences in halo mass estimation may also play a role. Using the eROSITA sample, which combines optically and X-ray selected galaxy groups and clusters, we find strong consistency with the ACT kSZ observations, and thus base our main results on these data sets.

Using the best-fit parameters from the joint kSZ and eROSITA analysis, we compute the resulting 3D gas profiles across two characteristic mass regimes. For low-mass, CMASS-like halos, the inferred gas profiles show strong central suppression and enhanced outskirts, consistent with a scenario of powerful feedback processes redistributing gas. At the high-mass end, our model accurately reproduces both the gas density and pressure profiles inferred from the X-COP sample, despite not being fitted to these measurements, highlighting the model's physical consistency and predictive power across halo masses.

We also compute the suppression of the matter power spectrum predicted by the best-fit model. The resulting suppression exceeds the percent level at \( k \gtrsim 0.3\,h\,\mathrm{Mpc}^{-1} \), reaching 20–25\% at \( k = 5\,h\,\mathrm{Mpc}^{-1} \), in line with strong-feedback predictions from the FLAMINGO \( f_{\mathrm{gas}} \)--8\(\sigma\) run. This result clearly deviates from the fiducial FLAMINGO \( m8 \) simulation, and points towards strong baryonic feedback effects required to jointly describe the combination of gas fractions from eROSITA and kSZ observations from ACT. This is in line with previous work (e.g.~\cite{Bigwood2024, McCarthy_kSZ2024}) and it will be interesting to compare it with recent and upcoming weak lensing results.

Taken together, our results show that the BFC framework provides a powerful, self-consistent model of baryonic feedback across mass scales and observables, including gas fractions, SZ effects, and matter clustering. Its predictive accuracy and physical interpretability make it highly promising for application to additional cosmological observables such as thermal SZ or X-ray observations, as well as simulation-based inference analyses. As upcoming surveys like Euclid, LSST, and Roman push into the nonlinear regime with unprecedented precision, robust modeling of baryonic effects, such as that provided by BFC, will be essential for extracting unbiased cosmological information.

\begin{acknowledgments}
We sincerely thank Vittorio Ghirardini for sharing the X-COP data, and Paola Popesso, and Emmanuel Schaan for providing access to their data. We also thank Tilman Tröster and Alexandre Refregier for valuable feedback and discussions. We as well acknowledge the access to the Marvin cluster of the University of Bonn. AN acknowledges support from the European Research Council (ERC) under the European Union’s Horizon 2020 research and innovation program with Grant agreement No. 101163128. SKG acknowledges theis supported by NWO (grant nonumber OCENW.M.22.307) and Olle Engkvist Stiftelse (grant no. 232-0238). JB acknowledges that this work was partially supported by the Swiss National Science Foundation under grant no. 200021\_192243. This research used DiRAC@Durham, operated by the Institute for Computational Cosmology and supported by the STFC DiRAC HPC Facility. Funding was provided via STFC capital grants ST/K00042X/1, ST/P002293/1, ST/R002371/1, and ST/S002502/1, as well as Durham University and STFC operations grant ST/R000832/1.
This work made use of several open-source tools, including \textsc{NumPy}~\cite{Harris_numpy_2020}, \textsc{SciPy}~\cite{Virtanen_scipy_2020}, \textsc{matplotlib}~\cite{Hunter_matplotlib_2007}, and \textsc{ChainConsumer}~\cite{Hinton_ChainConsumer_2016}.
\end{acknowledgments}

\newpage
\appendix
\section{Two-halo term and halo exclusion}\label{appendix: two-halo}

In this appendix, we derive the approximate two-halo term used in our modeling and discuss the influence of halo exclusion. The approximation follows the standard halo model formalism~\cite{Cooray_Sheth_halo_model_2002}, where the two-halo contribution in real space is obtained by Fourier transforming the two-halo power spectrum.

In the large-scale limit ($k \to 0$), the Fourier-transformed one-halo profile becomes:
\begin{equation}
    u(k|m) = \int_0^\infty dr\, 4\pi r^2 \frac{\sin(kr)}{kr} \zeta_{1\mathrm{h}}(r|m) \longrightarrow M_{\mathrm{gas}}(m)\;.
\end{equation}
Inserting this into Eq.~\eqref{eq: P_2h} and taking the same limit, we find:
\begin{equation}
    P_{\mathrm{2h}}(k \rightarrow 0) \approx b(m) P_\mathrm{lin}(k) \int_0^\infty \mathrm{d}m'\, \frac{\mathrm{d}n}{\mathrm{d}m'}\, b(m') M_{\mathrm{gas}}(m') \approx b(m) P_\mathrm{lin}(k) \bar{\rho}_\mathrm{gas}\;.
\end{equation}
This expresses the idea that the contribution from surrounding halos can be approximated as a uniform background with mean gas density $\bar{\rho}_\mathrm{gas} = f_\mathrm{gas} \cdot \rho_{\mathrm{m,crit}}$, modulated by the linear bias of the central halo.
 
Transforming back to real space, the approximate two-halo term becomes:
\begin{equation}
    \rho_\mathrm{2h}(r|m) = \bar{\zeta}_\mathrm{2h} \cdot \left(1 + b(m) \cdot \zeta_\mathrm{lin}(r) \right)\;,
\end{equation}
where $\zeta_\mathrm{lin}(r)$ is the linear correlation function. To ensure physically realistic behavior at large radii, we additionally include a constant background term \( \overline{\zeta}_\mathrm{2h} \), representing the mean cosmic density of the quantity being modeled. Without this term, the profile would asymptotically approach zero at large scales, which is inconsistent with the presence of a uniform cosmic background.
Finally, by adding the expression of halo-exclusion (shown in~\ref{subsec: 2-halo term}), we get the final form of the approximative two-halo term:
\begin{equation}
    \rho_\mathrm{2h}(r|m) = \Theta_\mathrm{e}(r|m) \cdot \bar{\zeta}_\mathrm{2h} \cdot \left(1 + b(m) \cdot \zeta_\mathrm{lin}(r) \right)\;.
\end{equation}

\begin{figure*}
\begin{center}
\includegraphics[width=0.49\textwidth]{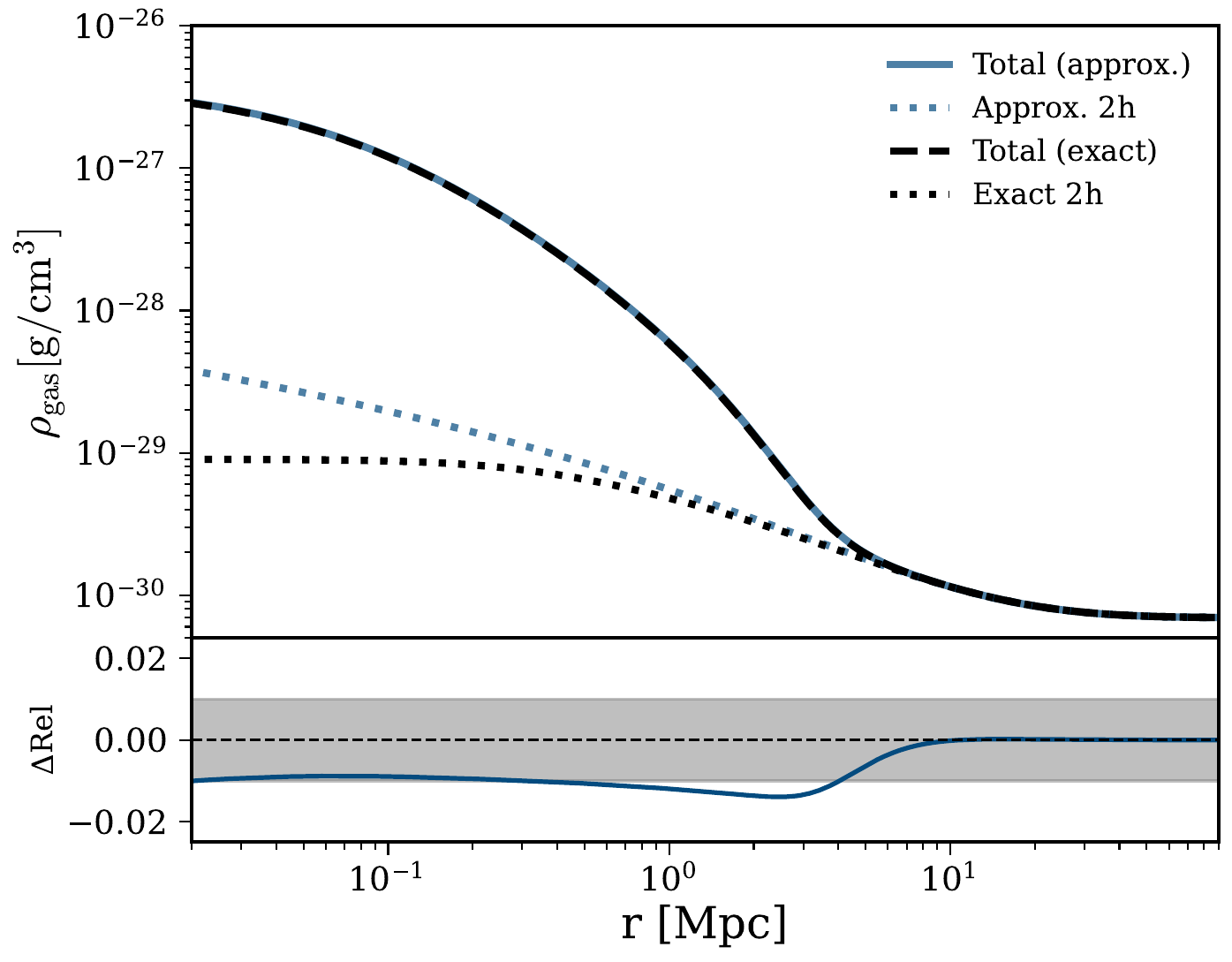}
\includegraphics[width=0.49\textwidth]{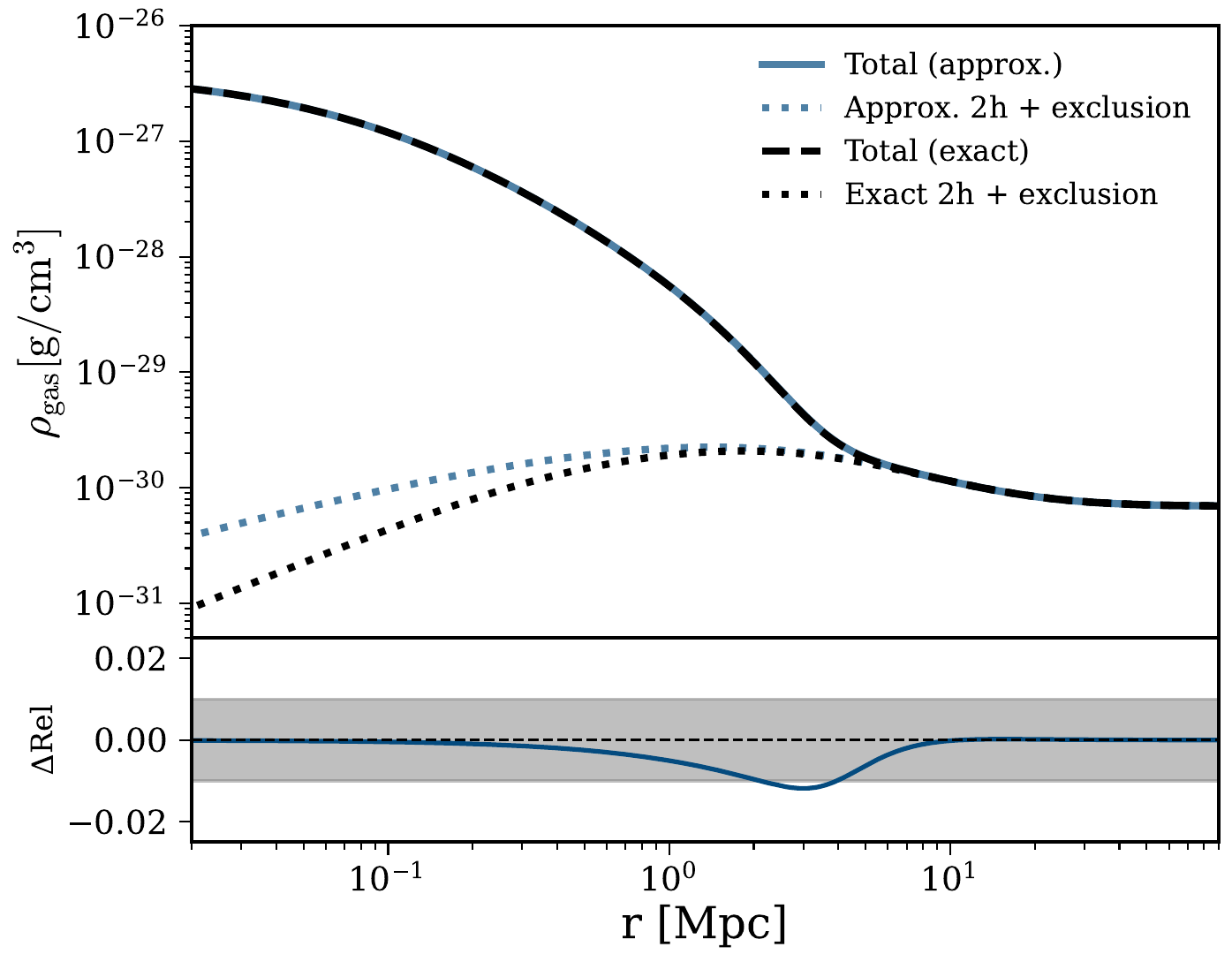}
\caption{\textbf{Comparison between the exact and approximate two-halo term, with and without halo exclusion.} \textit{Left:} Comparison of the exact and approximate two-halo terms without applying the halo-exclusion function. \textit{Right:} Same comparison, but with the exclusion term included in the approximate model. The exact two-halo term is computed via the Fourier transform of the two-halo power spectrum (Eq.~\eqref{eq: zeta_2h}), while the approximate version is given by Eq.~\eqref{eq: approx. 2-halo term}, which includes a background term and, in the right-handed-side panel, the exponential exclusion function. Both panels also display the relative difference in a sub-panel below, showing deviations below 2\% across the relevant radial range.} \label{fig: two_halo comparison}
\end{center}
\end{figure*}

In Figure~\ref{fig: two_halo comparison}, we compare the exact two-halo term (from full Fourier integration), the approximate expression without halo exclusion, and the same expression with the exclusion correction applied. We find that all models agree to better than 2\% across the relevant radial range. The effect of halo exclusion is limited to radii $r \lesssim r_\mathrm{vir}$ and has a negligible impact on the scales most relevant for our modeling.

\section{Parameter dependence of kSZ and $f_\mathrm{gas,500}$}
\begin{figure}[htbp]
\centering
\includegraphics[width=\linewidth]{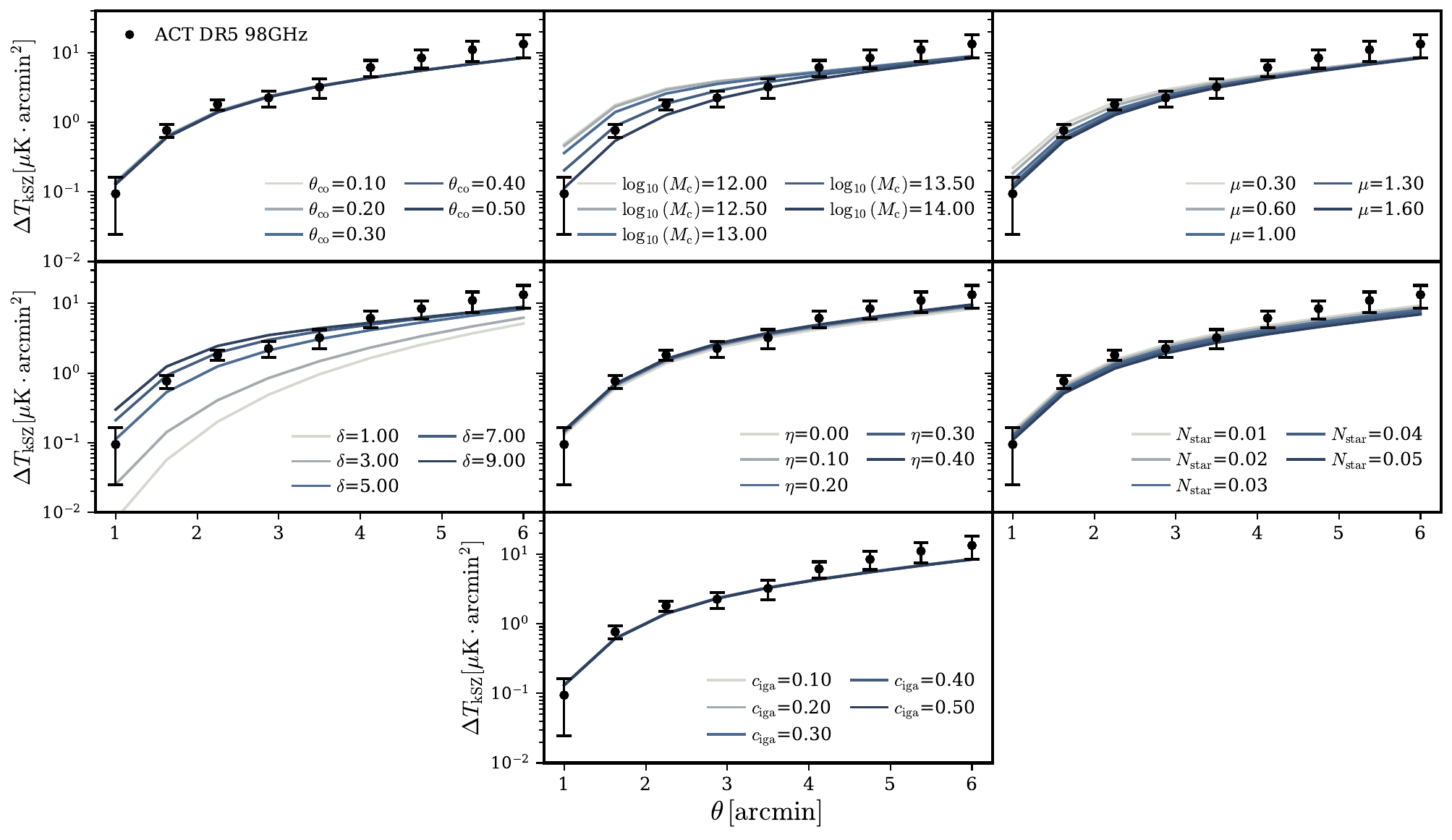}
\caption{Parameter dependence of the kSZ signal.
Each panel shows how varying a single model parameter affects the predicted observable, while all other parameters are held fixed at their fiducial best-fit values from the joint kSZ + eROSITA fit. The color gradient from light blue to dark blue indicates increasing values of the varied parameter. 
} \label{fig: ksz_params}
\end{figure}

\begin{figure}[htbp]
    \centering
    \includegraphics[width=\linewidth]{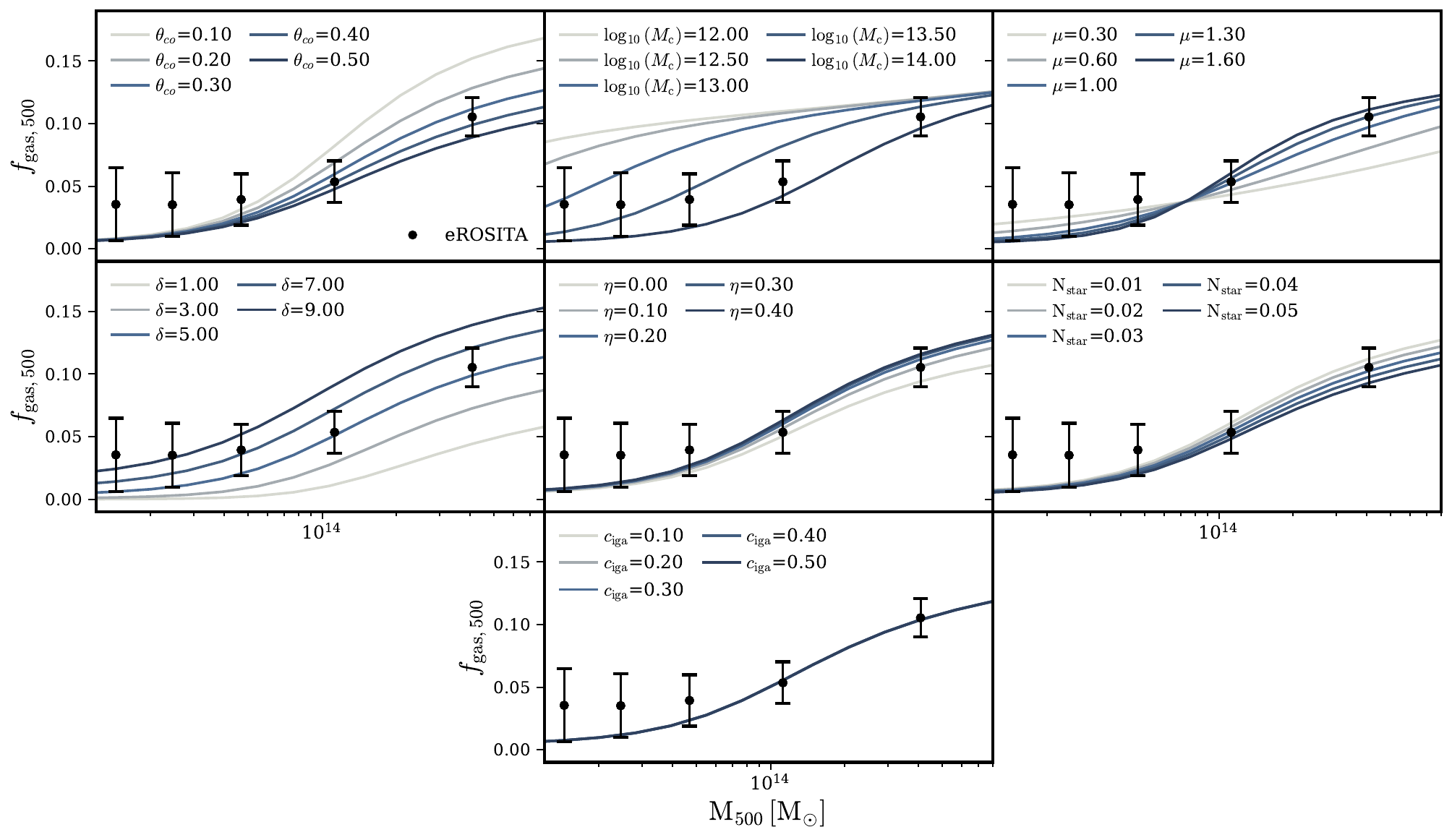}
    \caption{Similar to Figure~\ref{fig: ksz_params} but for the gas fractions.}
    \label{fig: fgas_params}
\end{figure}

Since the influence of the specific used parameters might not be directly obvious, we show the parameter dependence of the kSZ and gas fractions $f_\mathrm{gas,500}$ in Figures~\ref{fig: ksz_params} and~\ref{fig: fgas_params}, respectively. The relevant definitions for the gas density profile and baryonic fractions are provided in Equations~\eqref{eq: rho_hga equation}–\eqref{eq: fgas definition}. \\

\noindent\textbf{\boldsymbol{$\theta_\mathrm{co}$}:} Similar to $\alpha$, $\theta_\mathrm{co}$ also influences the inner part of the gas profile by defining how extended the core is. For the same reason as for $\alpha$ this has no influence on the kSZ. For the gas fractions we again see an effect. As the core radius becomes larger, we have less gas within $r_{500}$ and vice versa due to mass conservation. \\

\noindent\textbf{\boldsymbol{$\log_{10}(M_\mathrm{c})$}:} Defines the characteristic mass scale at which the transition in the $\beta(M)$ (see Eq. \eqref{eq: rho_hga beta}) slope occurs. This parameter influences how quickly the gas density profile steepens with increasing halo mass. For low masses (below $M_{\rm c}$), the profile is shallower; for high masses, it steepens. A lower $M_{\rm c}$ value leads to steeper gas profiles at fixed halo mass. This results in more gas concentrated within the inner regions of the halo, increasing both the gas fraction within $r_{500}$ and the amplitude of the kSZ signal.\\

\noindent\textbf{\boldsymbol{$\mu$}:} Controls the sharpness of the transition in the slope $\beta(M)$. A larger $\mu$ results in a more abrupt change around the transition mass $M_{\rm c}$, while smaller values yield a smoother variation. This affects both the shape of the gas profile and the mass dependence of the gas fraction and kSZ signal. A sharper transition can lead to more pronounced suppression or enhancement of gas at certain masses.\\

\noindent\textbf{\boldsymbol{$\delta$}:} Regulates how sharply the profile is truncated at large radii. A small $\delta$ results in a flatter profile that extends to large radii, redistributing more gas to the outskirts and lowering the central density. This leads to a suppressed kSZ signal in the center and lower gas fractions within $r_{500}$, since more mass lies beyond this radius. Conversely, large $\delta$ values steepen the outer profile, concentrating more gas inward and increasing both the kSZ signal and the gas fraction.\\

\noindent\textbf{\boldsymbol{$\eta$}:}
Controls the high-mass slope of the stellar fraction function. Larger values of $\eta$ reduce the stellar content in massive halos, thereby increasing the remaining hot gas fraction. The effect on the kSZ is moderate and comes from the fact that the total hot gas content changes, especially in the most massive halos. \\

\noindent\textbf{\boldsymbol{$N_{\rm star}$}:}
Sets the normalization of the stellar fraction, thereby indirectly reducing the available baryonic budget for gas. A higher $N_{\rm star}$ increases the stellar mass fraction, reducing the total hot gas fraction. This mainly affects the gas fraction constraints, especially at intermediate to high halo masses. The effect on the kSZ is weaker, but still present due to the reduced gas content. \\

\noindent\textbf{\boldsymbol{$c_{\rm iga}$}:}
Defines the fraction of central (cold) gas relative to the central galaxy stellar component. Increasing $c_{\rm iga}$ boosts the amount of cold, inner gas assumed to be unavailable for the hot phase, reducing the remaining hot gas fraction. This has no significant effect on the kSZ and the gas fractions, which is only sensitive to the ionized hot gas.

\section{Parameter reduction}\label{appendix: param reduc}\label{appendix: param reduc observations}
\begin{figure}[htbp]
\centering
\includegraphics[width=\linewidth]{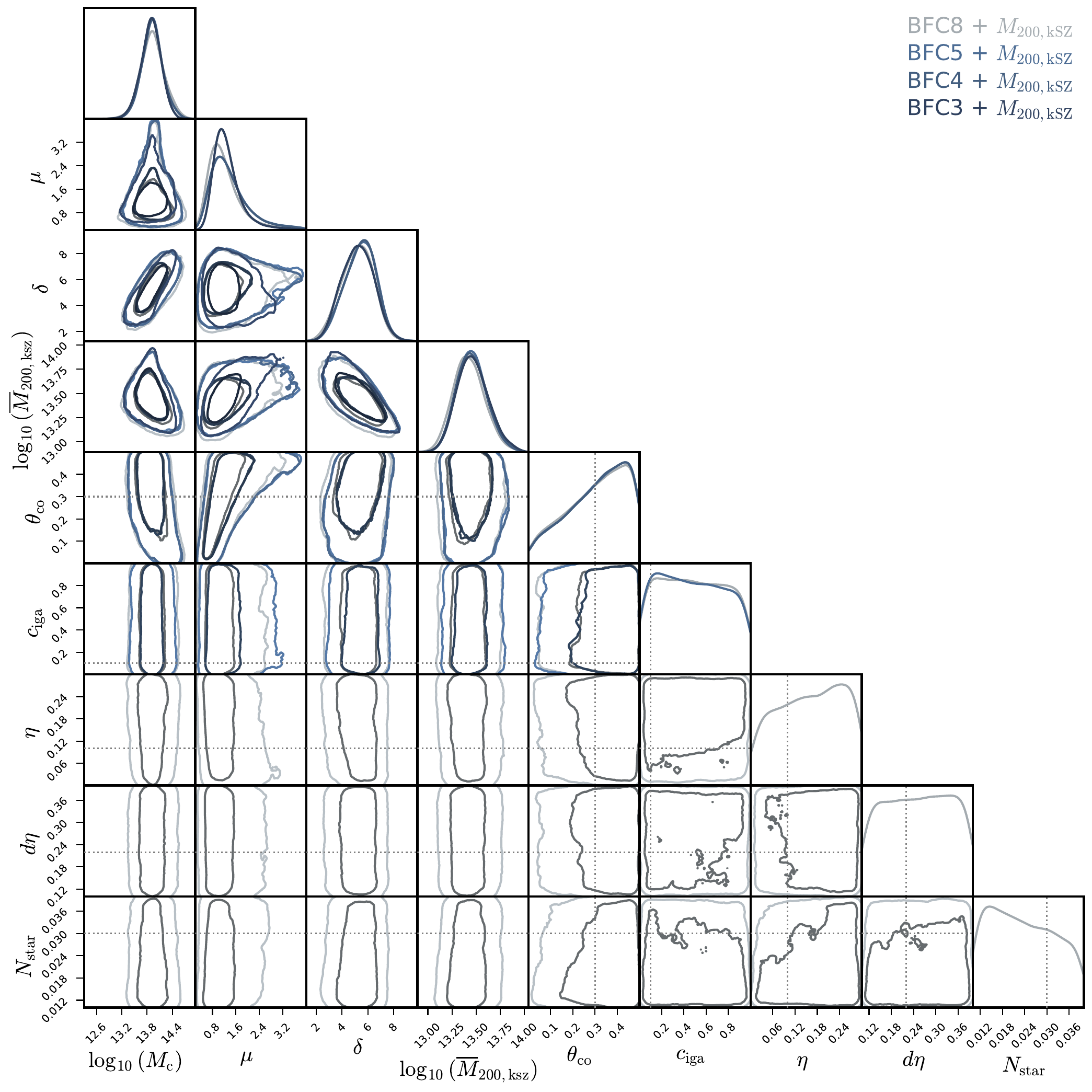}
\caption{\textbf{Parameter reduction from the joint kSZ + eROSITA gas fraction fits}.\\
The corner plot illustrates the stepwise parameter reduction applied in the joint fit to the kSZ signal and X-ray gas fractions. Starting from the full set of nine free parameters (light blue, $\mathrm{BFC8}+M_{200,\mathrm{kSZ}}$), we progressively fix parameters that are either unconstrained or have negligible impact on the observables: first the stellar parameters $\eta$, $d\eta$, and $N_{\mathrm{star}}$, followed by the cold gas fraction parameter $c_{\mathrm{iga}}$, and finally the transition scale $\theta_{\mathrm{co}}$. The final model (dark blue, $\mathrm{BFC3}+M_{200,\mathrm{kSZ}}$) retains only $M_\mathrm{c}$, $\mu$, $\delta$, and the effective halo mass. Dotted lines mark the fixed values, and the contours show how the posteriors evolve with each reduction step. \label{fig: eksz param reduction}}
\end{figure}

In this section, we describe the parameter reduction procedure used for the joint fit to the kSZ signal and X-ray gas fractions. The goal of this reduction is to simplify the model while retaining its ability to describe the data accurately.\\
We begin with a total of nine free parameters ($\mathrm{BFC8}$ + $M_{200,\rm kSZ}$): eight from the baryonification model, five governing the gas distribution ($M_\mathrm{c}$, $\mu$, $\delta$, $\theta_\mathrm{co}$, and $c_\mathrm{iga}$) and three controlling the stellar density profile ($\eta$, $d\eta$, and $N_{\mathrm{star}}$), along with the effective halo mass $M_{200,\mathrm{kSZ}}$ for the kSZ measurement. The parameter reduction proceeds in three steps, illustrated in Fig.~\ref{fig: eksz param reduction}, which shows the impact of each step on the resulting model constraints. \\
In the first step, we fix the stellar parameters $\eta$, $d\eta$, and $N_{\mathrm{star}}$. These parameters influence the gas fraction only indirectly by shaping the stellar mass component and thereby affecting the baryon budget. Since neither the kSZ signal nor the X-ray gas fractions directly constrain the stellar distribution, these parameters remain largely unconstrained and can be fixed without affecting the fit quality.\\
Next, we fix the cold gas fraction parameter $c_{\mathrm{iga}}$. This parameter affects the gas density only in the innermost regions of halos, which are not probed by the cosmological observables used in this analysis. Consequently, its impact on the inferred gas fractions and the kSZ signal is negligible.\\
Finally, we fix the transition parameter $\theta_{\mathrm{co}}$, which controls the location of the steepening in the gas profile. Similar to $c_{\mathrm{iga}}$, this parameter primarily affects the central regions of halos and has limited influence on the large-scale gas distribution relevant for our observables.\\
After these reductions, we are left with four free parameters ($\mathrm{BFC3}$ + $M_{200,\rm kSZ}$): $M_\mathrm{c}$, $\mu$, and $\delta$, which shape the gas profile, and the effective halo mass $M_{200,\mathrm{kSZ}}$. These remaining parameters are well constrained by the joint dataset and are not significantly affected by fixing the others.

\section{Posterior Distributions from Joint Fits to kSZ and X-ray Gas Fractions}\label{appendix: posterior comparison}
\begin{figure}
    \centering
    \includegraphics[width=0.49\linewidth]{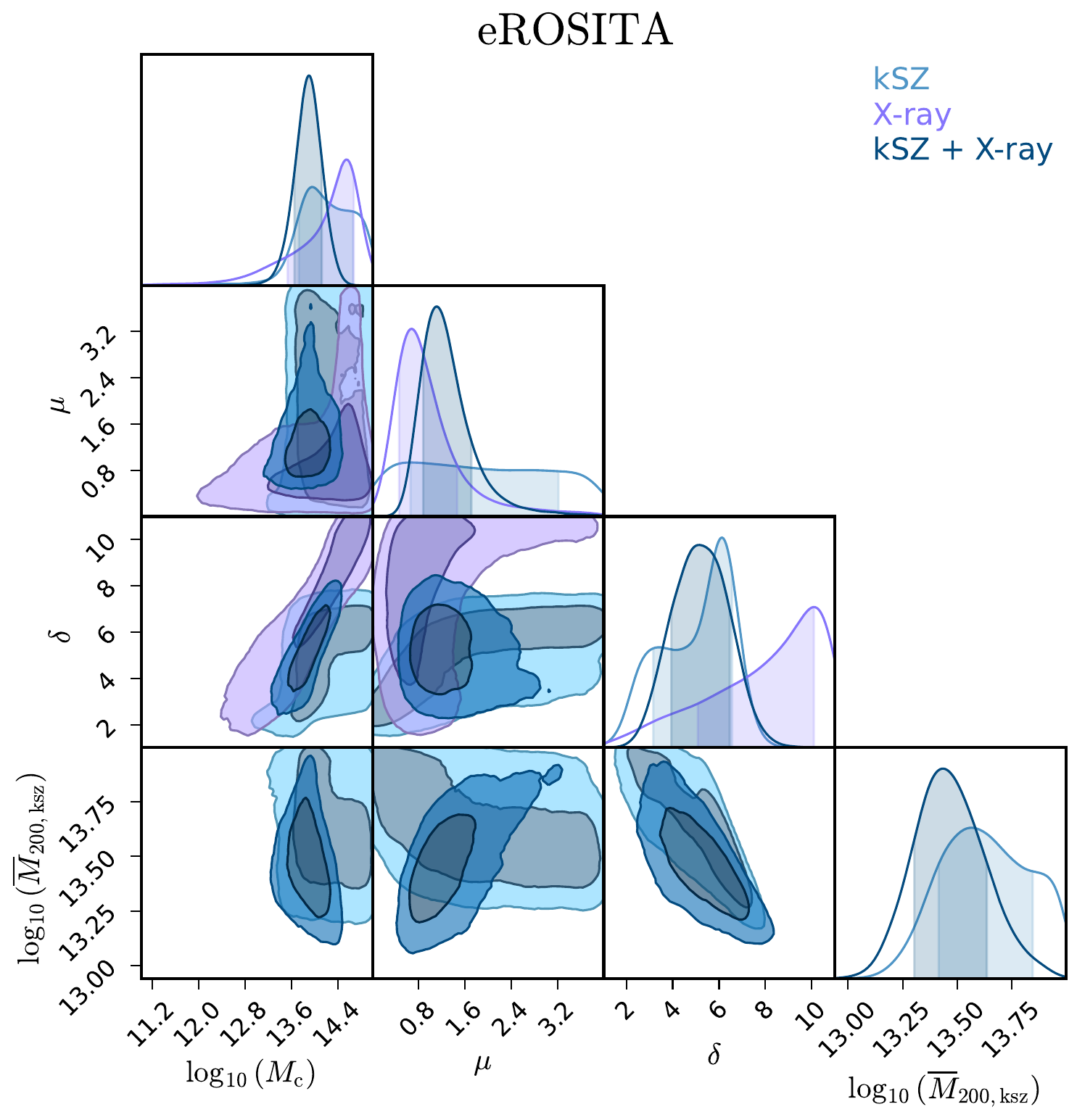}
    \includegraphics[width=0.49\linewidth]{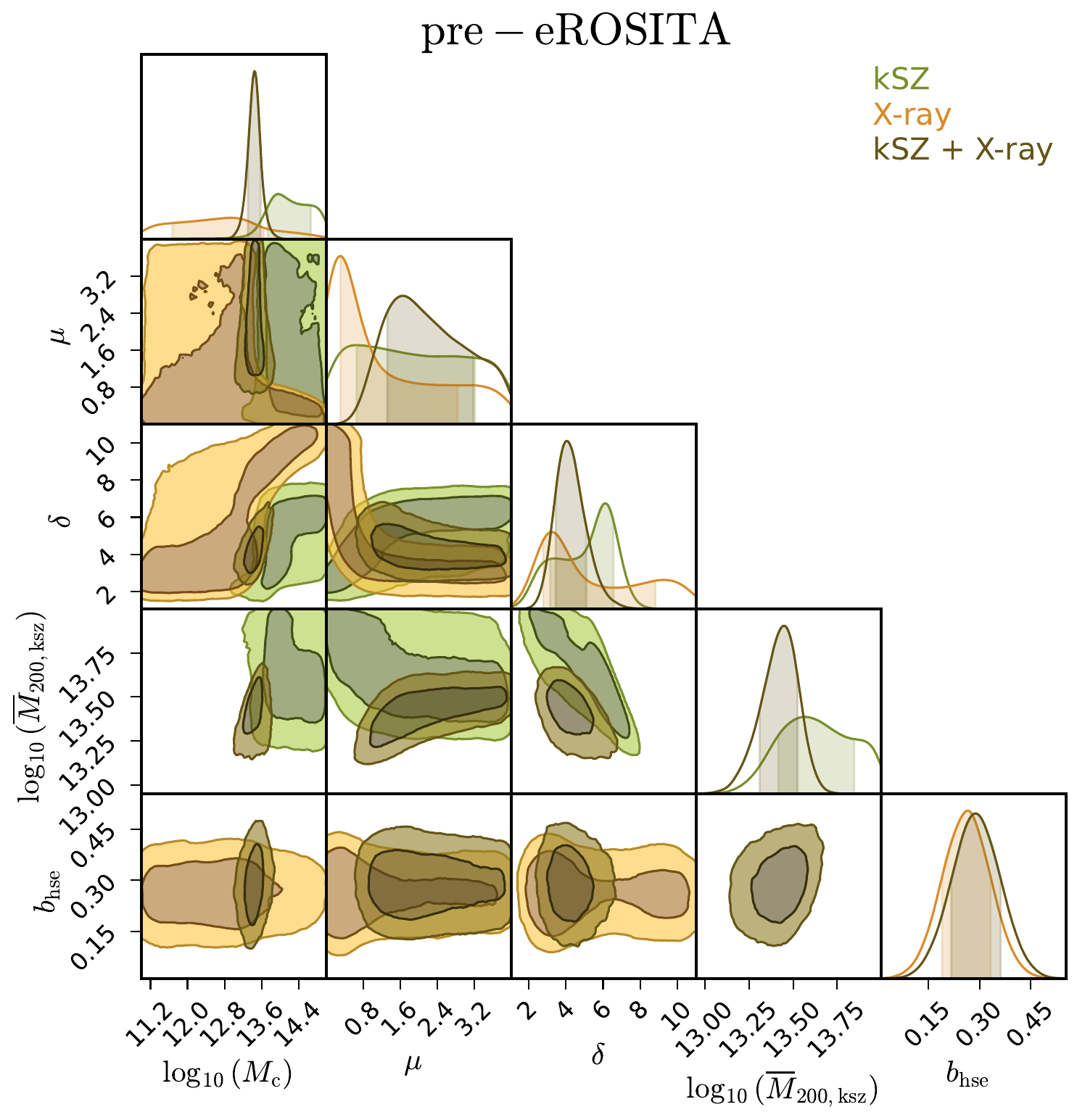}
    \caption{\textbf{Posterior distributions from the joint kSZ + X-ray fits.} Left: Posterior distributions from the kSZ + eROSITA fit. Right: Posterior distributions from the kSZ + pre-eROSITA fit. Both panels show the individual fits and the resulting joint posterior.}
    \label{fig: corner 3p eksz comparison}
\end{figure}

Figure~\ref{fig: corner 3p eksz comparison} presents the marginalized posterior distributions of the model parameters for the joint fits to kSZ and X-ray gas fraction data. The left-hand panel shows the results for the combination of kSZ with the eROSITA gas fractions, including the individual constraints from each probe and their joint posterior. The posteriors of all free parameters are largely consistent between the two probes, with significant overlap in their distributions. A closer inspection reveals that $\mu$ is primarily constrained by the gas fraction data, while $\delta$ is more tightly determined by the kSZ measurements, as indicated by their respective shifts in the individual fits.

The right-hand panel shows the same analysis using the pre-eROSITA gas fractions instead. In this case, the individual fits are less compatible. Notably, the characteristic mass scale $M_\mathrm{c}$ is favored to be significantly lower by the gas fraction data and higher by the kSZ data. The joint posterior settles in between the two, reflecting a compromise between the conflicting constraints. A similar tension appears in the $\mu$ parameter. Overall, while a joint fit is still possible, it requires more extreme parameter values, and the discrepancy between the individual datasets suggests limited consistency in the underlying constraints.

\section{pre-eROSITA power spectrum suppression} \label{appendix: pre-erosita pk sup}
\begin{figure}
    \centering
    \includegraphics[width=0.75\linewidth]{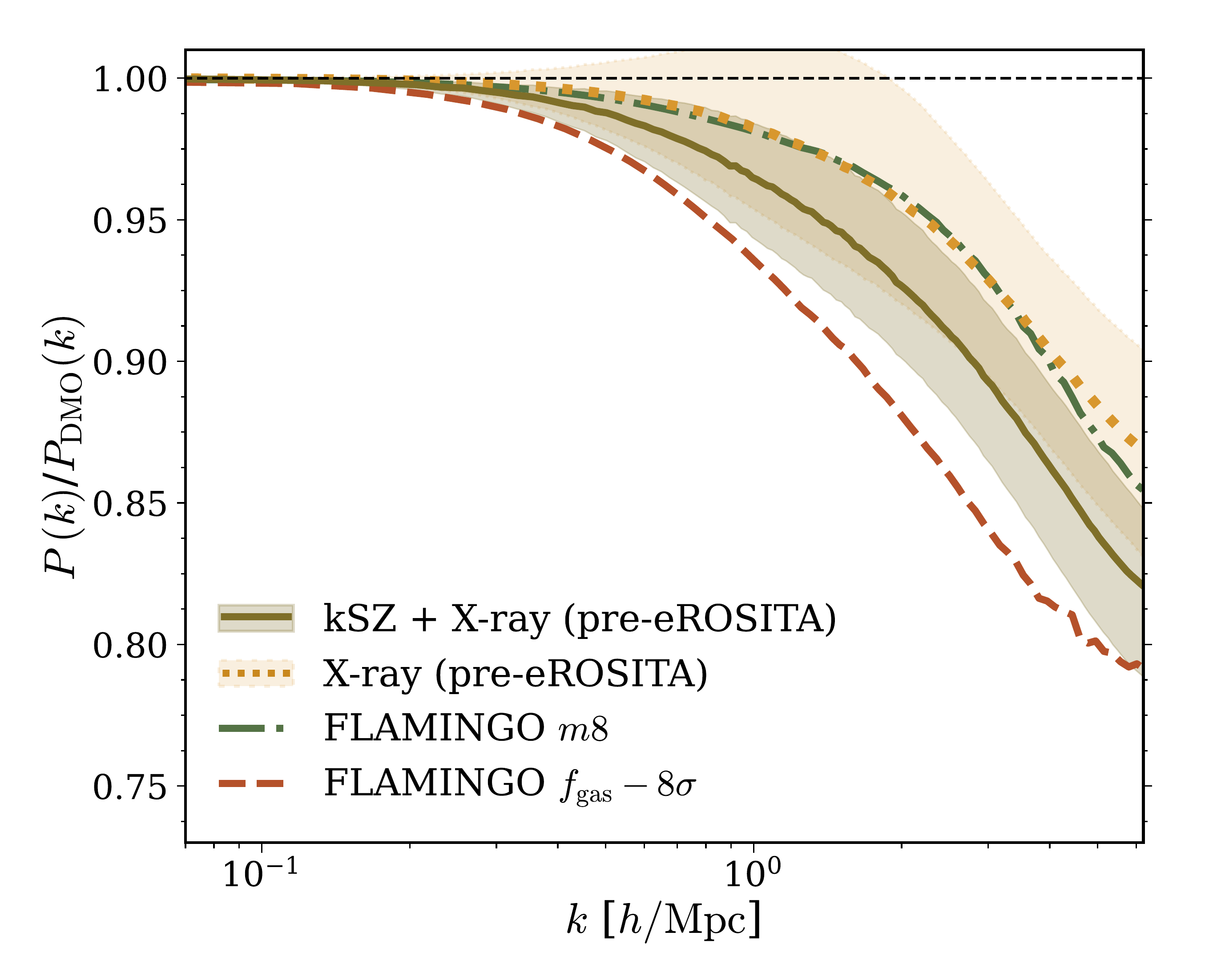}
    \caption{Matter power spectrum suppression from the joint fit using pre-eROSITA gas fractions. The solid brown line shows the predicted suppression using the best-fit parameters from the joint fit to kSZ and pre-eROSITA gas fraction data, with the shaded region indicating the 68\% confidence interval. For comparison, the green and red curves show the suppression from the $m8$ and $f_{\mathrm{gas}}$–8$\sigma$ FLAMINGO simulations, respectively. We also show the result from fitting the pre-eROSITA gas fractions alone (dotted golden line). The suppression lies between the two FLAMINGO models, and aligns closely with the fiducial simulation when only fitting gas fractions, consistent with the fact that the FLAMINGO fiducial model was calibrated against this earlier dataset.
}

    \label{fig: pre-erosita pk sup}
\end{figure}

In Figure~\ref{fig: pre-erosita pk sup}, we show the suppression of the matter power spectrum predicted by the best-fit parameters obtained from the joint fit to the kSZ signal and pre-eROSITA gas fractions. The result is shown as the solid brown line, with the shaded band indicating the 68\% confidence interval. The best-fit prediction lies between the $m8$ FLAMINGO simulation and the strong feedback $f_{\mathrm{gas}}$–8$\sigma$ model. All three curves agree well on large scales ($k < 0.2,h,\mathrm{Mpc}^{-1}$), where baryonic effects are subdominant. We also show the result when fitting only the pre-eROSITA gas fractions (dotted golden line). In this case, the predicted suppression closely matches the $m8$ FLAMINGO model, which is reassuring given that FLAMINGO was calibrated against these older gas fraction measurements.

As in the eROSITA-based analysis, we find that adding the kSZ data enhances the power suppression. This is again driven by the kSZ preference for a flatter profile, consistent with stronger feedback. Compared to the results based on eROSITA gas fractions, the joint kSZ + pre-eROSITA fit predicts less suppression overall. This is in line with the comparison of gas fractions themselves (compare gas fractions in Fig.~\ref{fig: eksz_fit} and~\ref{fig: fgas_joint_fit_sksz}), where the pre-eROSITA data favors higher gas fractions than the updated eROSITA measurements. Taken together, these results reinforce the conclusion that the eROSITA gas fractions are more consistent with a strong feedback scenario, while the pre-eROSITA results align better with the $m8$ feedback strength in FLAMINGO.

\section{Influence of Satellites}\label{appendix: satellite discussion}
In this section, we assess the impact of satellite galaxies on the kSZ modeling and the resulting constraints on baryonic feedback. Our fiducial analysis assumes that the observed kSZ signal originates solely from central galaxies. However, satellite galaxies also contribute to the overall signal, and their omission could potentially bias the inferred parameters.

To test the significance of this effect, we follow the approach presented in Ref.~\cite{McCarthy_kSZ2024}, where the contribution of satellites to the kSZ signal is quantified as a function of the minimum satellite mass. Specifically, their Figure 4 provides the ratio $T_{\rm kSZ,\,cent} / T_{\rm kSZ,\,cent+sats}$ for different satellite mass thresholds. We adopt their suggested cutoff of $M_{\rm min,\,sat} = 10^{11.2}\,\mathrm{M}_\odot/h$ and rescale our modeled kSZ signal accordingly, thus approximately incorporating the satellite contribution.

\begin{figure}
    \centering
    \includegraphics[width=0.99\linewidth]{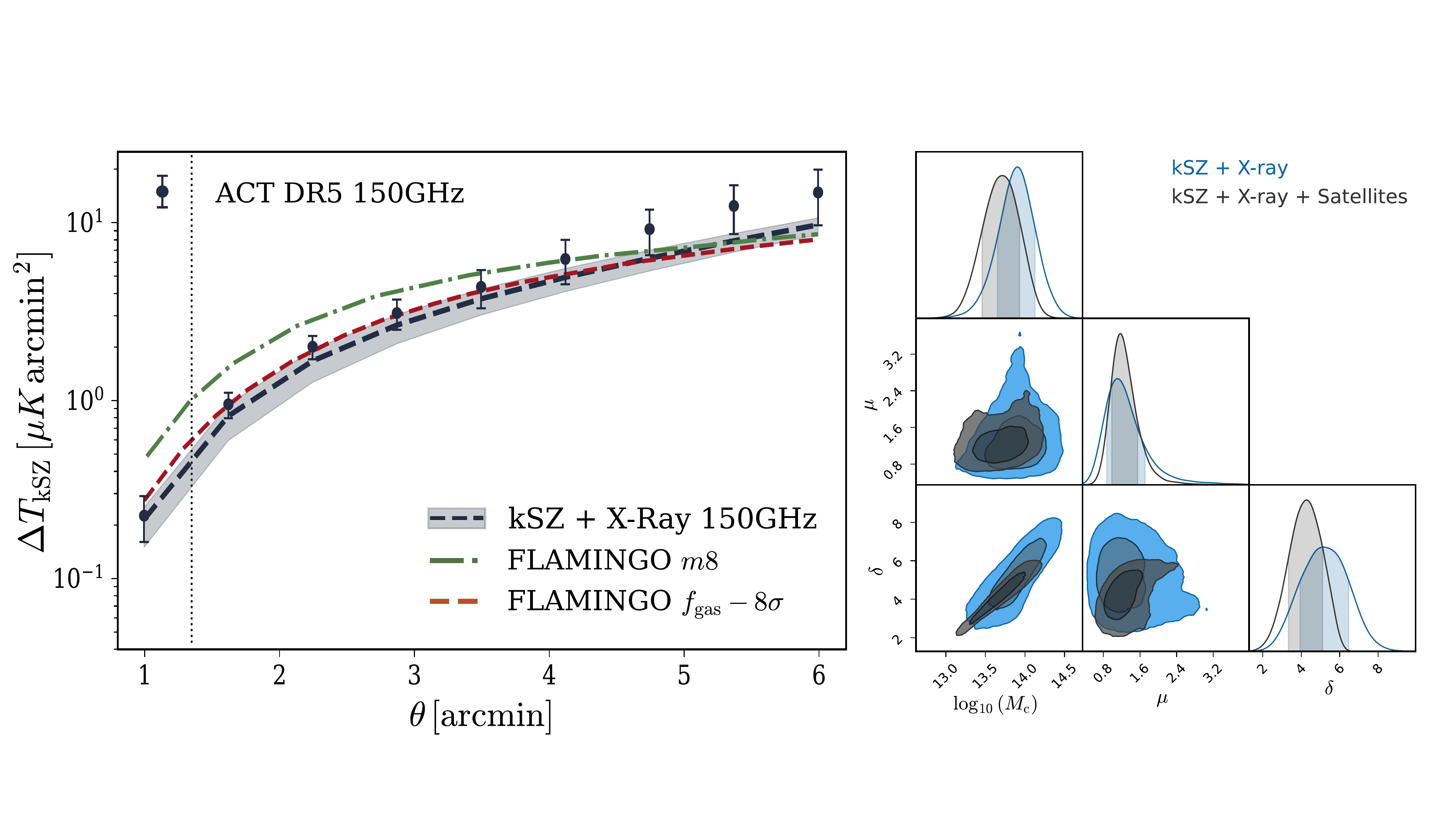}\\
    \includegraphics[width=0.7\linewidth]{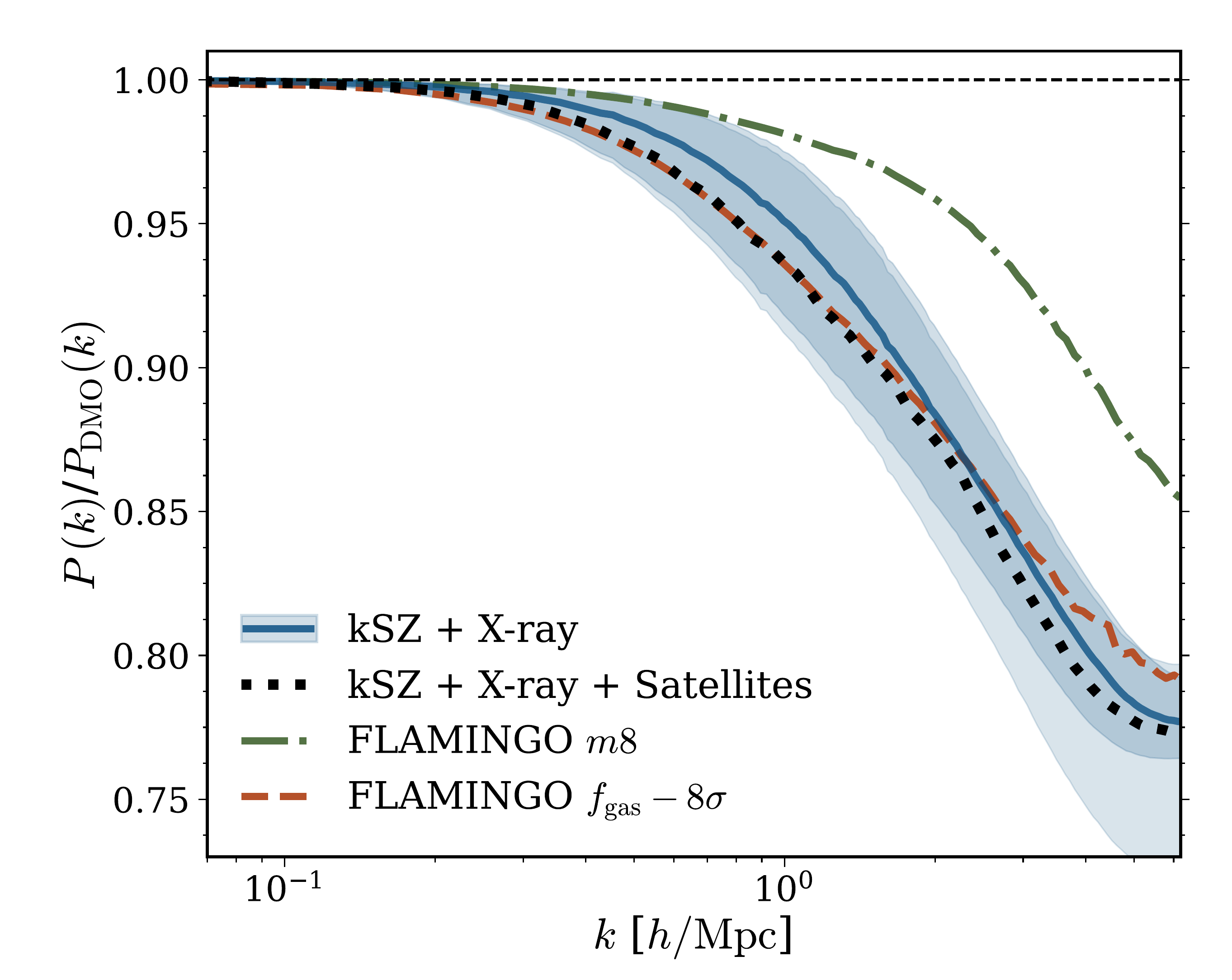}
    \caption{
\textbf{Impact of satellite galaxies on the kSZ modeling.} \textit{Top left:} Predicted kSZ signal from the best-fit model including satellite contributions (dashed black), compared to the kSZ signal measured in the FLAMINGO simulations for the $f_\mathrm{gas}-8\sigma$ scenario (dashed red) and the fiducial $m8$ run (dash-dotted green). \textit{Top right:} Posterior contours (68\% and 95\%) of the key gas model parameters, showing only minor shifts between the fiducial fit (blue, centrals only) and the satellite-corrected fit (black). \textit{Bottom center:} Suppression of the matter power spectrum predicted from the fiducial best-fit model (solid line with 68\% confidence band), compared to the satellite-included prediction (dotted line), which remains well within the fiducial confidence interval.
}
\label{fig: pk_suppression_satellite_test}
\end{figure}

Using this modified signal, we rerun the full MCMC chain for the joint fit to the kSZ and eROSITA gas fractions. We then compute the corresponding matter power spectrum suppression from the new best-fit parameters. The result is shown in Figure~\ref{fig: pk_suppression_satellite_test} (lower center panel), where we compare the fiducial suppression (solid blue line with 68\% confidence interval) to the case including satellites (dotted black line). We find that the new result lies well within the 68\% confidence region of the fiducial fit, indicating that the omission of satellites does not significantly affect the level of suppression inferred from current data.

In addition, we compare the posterior distributions of the key baryonic parameters in Figure~\ref{fig: pk_suppression_satellite_test} (right-handed-side panel). The posteriors shift only slightly, remaining consistent with the fiducial constraints within statistical uncertainties. This further confirms that the modeling choice to neglect satellites is justified for the current signal-to-noise level. However, for future high-precision kSZ measurements, explicit modeling of the satellite contribution will be essential to avoid systematic biases.

\nocite{*}
\bibliographystyle{JHEP}
\bibliography{biblio.bib}

\end{document}